\documentclass[aps,prx,twocolumn,nofootinbib]{revtex4-2}
\usepackage{amsmath}
\usepackage{amsfonts}
\usepackage{graphicx}
\usepackage{color}
\usepackage{dsfont}
\usepackage{verbatim}
\usepackage{mathtools}
\usepackage[normalem]{ulem}
\usepackage{comment}
\usepackage{stackrel}
\usepackage{bm}
\usepackage[pdftex]{hyperref}
\hypersetup{colorlinks = true, urlcolor = black, linkcolor = black, citecolor = black}

\definecolor{myblue}{rgb}{.93, .93, 1}

\newcommand{\bsub}{\begin{subequations}}
	\newcommand{\esub}{\end{subequations}}

\newcommand{\vex}[1]{\bm{\mathrm{#1}}}

\newcommand{\tr}{\mathsf{Tr}}

\newcommand{\ket}[1]{\left| {#1} \right\rangle}
\newcommand{\bra}[1]{\left\langle {#1} \right|}

\begin{document}
	
	\title{Scaling theory of intrinsic Kondo and Hund's rule interactions in magic-angle twisted bilayer graphene}
	
	\author{Yang-Zhi~Chou}\email{yzchou@umd.edu}
	\affiliation{Condensed Matter Theory Center and Joint Quantum Institute, Department of Physics, University of Maryland, College Park, Maryland 20742, USA}
	
	\author{Sankar~Das~Sarma}
	\affiliation{Condensed Matter Theory Center and Joint Quantum Institute, Department of Physics, University of Maryland, College Park, Maryland 20742, USA}
	
	\date{\today}

	\begin{abstract}
		Motivated by the recent studies of intrinsic local moments and Kondo-driven phases in magic-angle twisted bilayer graphene, we investigate the renormalization of Kondo coupling ($J_K$) and the competing Hund's rule interaction ($J$) in the low-energy limit. Specifically, we consider a surrogate single-impurity generalized Kondo model and employ the poor man's scaling approach.
		The scale-dependent $J_K$ and $J$ are derived analytically within the one-loop poor man's scaling approach, and the Kondo temperature ($T_K$) and the characteristic Hund's rule coupling ($J^*$, defined by the renormalized value of $J$ at some small finite energy scale) are estimated over a wide range of filling factors. We find that $T_K$ depends strongly on the filling factors as well as the value of $J_K$. Slightly doping away from integer fillings and/or increasing $J_K$ may substantially enhance $T_K$ in the parameter regime relevant to experiments. $J^*$ is always reduced from the bare value of $J$, but the filling factor dependence is not as significant as it is for $T_K$. Our results suggest that it is essential to incorporate the renormalization of $J_K$ and $J$ in the many-body calculations, and Kondo screening should occur for a wide range of fractional fillings in magic-angle twisted bilayer graphene, implying the existence of Kondo-driven correlated metallic phases.
		We also point out that the observation of distinct phases at integer fillings in different samples may be due to the variation of $J_K$ in addition to disorder and strain in the experiments.
	\end{abstract}
	
	\maketitle
	
	\section{Introduction}

	Studying moir\'e systems has become one of the central themes in condensed matter physics since the discovery of interaction-driven phases and superconductivity in magic-angle twisted bilayer graphene (MATBG) \cite{Cao2018_tbg1,Cao2018_tbg2}. Notably, the ability to tune the bandwidth in MATBG motivates numerous experiments aiming for correlation-induced phenomena \cite{Cao2018_tbg1,Cao2018_tbg2,Yankowitz2019,Kerelsky2019,Lu2019,Sharpe2019,Serlin2020,Jiang2019,Xie2019_spectroscopic,Polshyn2019,Cao2020PRL,Park2021flavour,Choi2019electronic,Zondiner2020cascade,Arora2020superconductivity,Wong2020cascade,Choi2021correlation}.
	While it is generally believed that interaction is the key to understanding the rich phases in MATBG \cite{Xu2018,Po2018PRX,Isobe2018PRX,Kang2019,Gonzalez2019,Lin2019,Wu2019_coupled-wire,Chou2019,Chen2020_correlated_PRB,Konig2020,Wu2020,Alavirad2020,Xie2020_Correlated_Insulator,Zhang2020_Correlated_ins,Bultinck2020,Cea2020,Huang2020,Liu2021_Th_correlated_ins,TBG_III,TBG_IV,TBG_V,Christos2020superconductivity,Chichinadze2020,Abouelkomsan2020,Soejima2020,Ledwith2020,Vafek2020,Kang2021,Chou2021,Wagner2022,Khalaf2021charged,Chen2021realization,TBGI,TBGII,Kwan2021,Thomson2021,Lewandowski2021,Wang2021,Shavit2021,Fischer2021}, a unified understanding encompassing all aspects is still missing. An outstanding problem from the theoretical perspective is the obstruction due to the $\mathcal{C}_{2z}\mathcal{T}$ symmetry preserving fragile topological minibands \cite{Song2019,Po2019,Ahn2019}, which suggests the absence of a lattice description for the two mini bands of MATBG. However, studying strongly correlated phenomena typically requires a well-defined lattice description. (In fact, strongly correlated phenomena are essentially a synonym for the physics of electron-electron interactions on a background lattice, as exemplified by, e.g.,  the paradigmatic Mott-Hubbard model.) Thus, it is desirable to derive an effective theory naturally compatible with the strong correlations, such as incorporating Hubbard interaction in the topological band picture.
	
	Recently, Song and Bernevig proposed a topological heavy fermion approach (THF) for magic-angle twisted bilayer graphene (MATBG) \cite{Song2022}. The idea is to reconstruct several low-energy bands of the Bistrizer-MacDonald (BM) model \cite{BM_model2011} by coupling a lattice of localized ``heavy'' $f$ fermions (akin to the zeroth pseudo-Landau levels at AA stacking registries \cite{Liu2019_Pseudo_LL,Shi2022HF_Rep}) and bands of topological ``itinerant conduction band'' $c$ fermions. The THF approach can overcome the topological obstruction in the low-energy minibands and allows for a reliable lattice model description \cite{Song2022,Cualuguaru2023tbg}, thus providing a unified strongly correlated description for MATBG. Within the THF formalism, the interacting MATBG Hamiltonian can be viewed as a generalized periodic Anderson model along with competing interactions \cite{Chou2022kondo,Hu2023kondo}, and the possibility of Kondo-driven and local-moment phases are explored in MATBG \cite{Chou2022kondo,Hu2023kondo,Hu2023symmetric,Zhou2023Kondo,Lau2023topological}. The THF ideas have been generalized to other moir\'e graphene systems \cite{Shi2022HF_Rep,Yu2023magic} and MATBG with a magnetic field \cite{Singh2023}.

	Among various possible strongly correlated scenarios, it is natural to investigate whether Kondo correlation is relevant to MATBG \cite{Chou2022kondo,Hu2023kondo,Hu2023symmetric,Zhou2023Kondo,Lau2023topological,Datta2023heavy,Huang2023evolution}. Using the THF formalism, a Kondo lattice description for MATBG can be derived straightforwardly in the local-moment regime \cite{Chou2022kondo,Hu2023kondo,Hu2023symmetric,Zhou2023Kondo}. Remarkably, hybridization between the $SU(8)$ local moments and itinerant $c$ fermions can realize a correlated topological (semi)metal within the Read-Newns mean-field approximation \cite{Chou2022kondo}.
	While Ref.~\cite{Chou2022kondo} propose a potential Kondo-driven correlated topological semimetal phase at the charge neutrality point (CNP), several works suggest a correlated insulating state at $\nu=0$ \cite{Hu2023symmetric,Zhou2023Kondo}. Besides the possible existence of various Kondo-driven phases, the local moments provide a natural explanation for the Pomeranchuk effect \cite{Hu2023kondo,Zhou2023Kondo}.
	
	In this work, we study the Kondo temperature and the renormalized Hund's rule interaction in MATBG using the THF formalism combined with the poor man's scaling approach \cite{Anderson1970poorman}. Our goal is to provide a simple semi-quantitative understanding of the filling-dependent Kondo correlation and the competing interaction in MATBG, which has not been investigated systematically. We incorporate the essential interaction effects in the minimal THF model and construct a surrogate single-impurity model incorporating the key features of the original Kondo lattice problem. We then derive analytical one-loop poor man's scaling flows for the Kondo coupling ($J_K$) and the Hund's rule interaction ($J$). Using the analytical flow equations, we obtain the Kondo temperature ($T_K$) and the characteristic Hund's rule interaction ($J^*$, defined by the renormalized $J$ at some small finite energy) as functions of filling factors associated with MATBG. While $T_K$ may be parametrically small near the integer fillings, doping slightly away from the integer filling can substantially enhance $T_K$. We also find that $T_K$ is exponentially sensitive to the value of $J_K$ for the parameter regime of interest. Thus, $T_K$ can be significantly boosted with a slightly larger $J_K$. Meanwhile, $J^*$ is always reduced from the bare value of $J$, but the filling factor dependence for $J^*$ is less significant than that for $T_K$. Our results indicate the importance of incorporating renormalization of $J_K$ and $J$ in the many-body calculations and suggest Kondo-driven correlated metals for a wide range of fractional fillings. In addition, the existence of Kondo-driven correlated (semi)metal phases of MATBG is very sensitive to the interaction strength near integer fillings. As such, the observation of distinct phases at integer fillings (e.g., insulator versus semimetal at the CNP) in different samples may very well be due to the variation of $J_K$ in addition to disorder and strain in the experiments.

	The rest of the article is organized as follows: In Sec.~\ref{Sec:Model}, we review the THF model and construct a surrogate single-impurity model for scaling calculations. Then, we derive the analytical poor man's scaling flows at the one-loop level in Sec.~\ref{Sec:PMS}. The analytical results with different parameters are analyzed systematically in Sec.~\ref{Sec:Results}. In Sec.~\ref{Sec:Discussion}, we discuss the implications of our results and possible generalization. Many important technical details are provided in the appendices. In Appendix~\ref{Sec:App:c_band_bare}, we provide the analytical expressions of the $c$-fermion band structures, wavefunctions, density of states, and hybridization functions. We do the same in Appendix~\ref{Sec:App:c_band_dyn_mass} except that the dynamical mass term is incorporated. The derivations of T-matrix contributions are given in Appendix~\ref{Sec:App:TMatrix}. In Appendix~\ref{Sec:App:Imp}, the flows of impurity potential scattering terms are summarized. We also discuss the poor man's scaling approach for an impurity Anderson model in Appendix~\ref{Sec:App:Anderson_PMS}.
	
	\section{Model}\label{Sec:Model}
	
	The low-energy mini bands of the BM model \cite{BM_model2011} for MATBG can be reconstructed through the THF model \cite{Song2022}, which describes a triangular lattice of localized $f$ fermions coupled to dispersing topological $c$ fermion bands. This reformulation provides a systematic way of constructing a lattice model in MATBG, which is crucial for studying physics driven by strong correlation. We first review the THF model and then introduce a surrogate single-impurity model, which can be used for studying the scaling properties of Kondo coupling and Hund's rule interaction.

	\subsection{Topological heavy fermion model}
	
	In the following, we review the THF model \cite{Song2022} proposed by Song and Bernevig in order to provide an essential background for our work. The main idea is to reconstruct the low-energy MATBG bands into hybridization between localized orbitals and delocalized topological bands. We also discuss the expression of the projected Coulomb interaction in the THF model.
	
	First, the localized states ($f$ fermions) have the $p_x\pm ip_y$-like Wannier orbitals \cite{Song2022} and are associated with the zeroth pseudo-Landau levels at AA registries \cite{Liu2019_Pseudo_LL,Shi2022HF_Rep}. These $f$-fermion orbitals form a triangular lattice with a lattice constant $a_M$, described by an exactly flat band
	\begin{align}\label{Eq:H_0_f}
		\hat{H}_{0,f}=-\mu_f\sum_{\vex{R}}\sum_{\alpha,\eta,s}f^{\dagger}_{\vex{R},\alpha,\eta,s}f_{\vex{R},\alpha,\eta,s},
	\end{align} 
where $\mu_f$ is the chemical potential of the $f$ fermions, and $f_{\vex{R},\alpha,\eta,s}$ is the $f$ fermion annihilation operator with valley $\eta$, spin $s$, orbital $\alpha$ ($\alpha=1,2$), and position $\vex{R}$.
 
Second, the delocalized topological bands ($c$ fermions) are described by a four-orbital $\vex{k}\cdot\vex{p}$ Hamiltonian given by
	\begin{align}\label{Eq:H_0_c}
		\hat{H}_{0,c}\!=\!\sum_{\eta,s,a,a'}\sum_{\vex{q}}c^{\dagger}_{\vex{q},a,\eta,s}\left[h^{(\eta)}_{aa'}(\vex{q})-\mu_c\delta_{aa'}\right]c_{\vex{q},a',\eta,s},
	\end{align} 
	where $\hat{h}^{(\eta)}(\vex{q})$ is a $4\times 4$ matrix, $\mu_c$ is the chemical potential of the $c$ fermion, $c_{\vex{q},a,\eta,s}$ is the $c$ fermion annihilation operator with valley $\eta$, spin $s$, orbital $a$ ($a=1,2,3,4$), and wavevector $\vex{q}$. 
	Note the $c$ fermion has a cutoff set by the microscopic graphene lattice constant instead of $a_M$. Thus, $\vex{q}$ is not restricted by the first moir\'e Brillouin zone defined by the triangular superlattice.
	
	Finally, the $c$ and $f$ fermions are coupled through a hybridization term given by \cite{Song2022}
	\begin{align}
		\label{Eq:H_0_cf}\hat{H}_{0,cf}\!\!=\frac{1}{\sqrt{\mathcal{N}_s}}\sum_{\vex{R}}\sum_{\eta,s,\alpha,a}\!\sum_{\vex{q}}\left[V^{(\eta)}_{\alpha a}(\vex{q})e^{i\vex{q}\cdot\vex{R}}f^{\dagger}_{\vex{R},\alpha,\eta,s}c_{\vex{q},a,\eta,s}\!+\!\text{H.c.}
		\right]\!,
	\end{align}
	where $\mathcal{N}_s$ is the number of sites in the moir\'e triangular lattice, and $V^{(\eta)}_{\alpha a}(\vex{q})$ is a $2\times 4$ hybridization matrix. 
	
	The matrices $\hat{h}^{(\eta)}$ and $\hat{V}^{(\eta)}$ are expressed by
		\begin{align}
		\hat{h}^{(\eta)}\!(\vex{q})\!=\!&\left[\!\begin{array}{cc}
			\hat{0}_{2\times 2} & v_*\!\left(\eta q_x\hat{\sigma}_0+iq_y\hat{\sigma}_z\right) \\[2mm]
			v_*\!\left(\eta q_x\hat{\sigma}_0-iq_y\hat{\sigma}_z\right) & M\hat{\sigma}_x		
		\end{array}\!\right],\\
		\hat{V}^{(\eta)}(\vex{q})=&e^{-\frac{|\vex{q}|^2\lambda^2}{2}}\!\!\left[\!\begin{array}{cc}
			\gamma\hat{\sigma}_0 + v_*'\!\left(\eta q_x\hat{\sigma}_x+q_y\hat{\sigma}_y\right), & \hat{0}_{2\times2}
		\end{array}
		\!\right].
	\end{align}
	In the above expressions, $\hat{\sigma}_{0}$ and $\hat{\sigma}_{\mu}$ represent the $2\times2$ identity operator and $\mu$-component of Pauli matrix respectively, $v_*=-4.303$ eV \AA, $M=3.697$ meV, $\gamma=-24.75$ meV, $v_*'=1.622$ eV \AA, $\lambda=0.3375a_M$, and $a_M=134.24$ \AA \cite{Song2022}. The parameters correspond to $w_0/w_1=0.8$, $w_1=110$ meV, and $\theta=1.05^{\circ}$ in the BM model \cite{Song2022}. The form factor $e^{-\frac{|\vex{q}|^2\lambda^2}{2}}$ in $\hat{V}^{(\eta)}$ is due to the finite width of the localized $f$ orbitals, which suppresses the contributions of large momenta.

	In addition to the single-particle bands, the THF model also contains interactions. The Coulomb interaction can be projected into the THF basis, and $\hat{H}_I=\hat{H}_U+\hat{H}_V+\hat{H}_W+\hat{H}_J$, where \cite{Song2022}
	\begin{align}
		\label{Eq:H_U}\hat{H}_U=&\frac{U}{2}\sum_{\vex{R}}:\hat{n}_{f,\vex{R}}::\hat{n}_{f,\vex{R}}:,\\
		\hat{H}_V=&\frac{1}{2}\int\limits_{\vex{r},\vex{r}'}:\hat{\rho}_c(\vex{r}):V(\vex{r}-\vex{r}'):\hat{\rho}_c(\vex{r}'):,\\
		\label{Eq:H_W}\hat{H}_W=&\Omega_0\sum_{\vex{R},a}W_a:\hat{n}_{f,\vex{R}}::\hat{\rho}_{c,a}(\vex{R}):,\\
		\nonumber\hat{H}_J
		=&-\frac{J}{\mathcal{N}_s}\sum_{\eta,\alpha,s_1,s_2}\sum_{\vex{R}}\sum_{\vex{q},\vex{q}'}e^{-i(\vex{q}'-\vex{q})\cdot\vex{R}}e^{-\frac{\lambda^2\left(|\vex{q}|^2+|\vex{q}'|^2\right)}{2}}\\
		\label{Eq:H_J_S}&\times\left[
		\begin{array}{c}
			:f^{\dagger}_{\vex{R},\alpha,\eta,s_1}f_{\vex{R},\alpha,\eta,s_2}::c^{\dagger}_{\vex{q}',\alpha+2,\eta,s_2}c_{\vex{q},\alpha+2,\eta,s_1}:\\[2mm]
			-f^{\dagger}_{\vex{R},\bar{\alpha},-\eta,s_1}f_{\vex{R},\alpha,\eta,s_2}c^{\dagger}_{\vex{q}',\alpha+2,\eta,s_2}c_{\vex{q},\bar{\alpha}+2,-\eta,s_1}
		\end{array}
		\right].
	\end{align}
	In the above expressions,  $:\hat{A}:$ denotes the normal order of $\hat{A}$, $\Omega_0$ is the area of a moir\'e unit cell, $\hat{n}_{f,\vex{R}}$ denotes the local number of $f$ fermions, $\hat{\rho}_{c,a}(\vex{r})$ denotes the local density of $c$ fermions with orbital $a$, $\hat{\rho}_c(\vex{r})=\sum_a\hat{\rho}_{c,a}(\vex{r})$, $\bar{1}=2$, and $\bar{2}=1$. The normal order is explicitly given by $:\hat{A}:\equiv A-\bra{G} A\ket{G}$, where
	\begin{subequations}
		\begin{align}
			\langle G|c^{\dagger}_{\vex{q},a,\eta,s}c_{\vex{q},a',\eta',s'}|G\rangle=&\frac{1}{2}\delta_{\vex{q},\vex{q}'}\delta_{a,a'}\delta_{\eta,\eta'}\delta_{s,s'},\\
			\langle G|f^{\dagger}_{\vex{R},\alpha,\eta,s}f_{\vex{R},\alpha',\eta',s'}|G\rangle=&\frac{1}{2}\delta_{\vex{R},\vex{R}'}\delta_{\alpha,\alpha'}\delta_{\eta,\eta'}\delta_{s,s'},\\
			\langle G|f^{\dagger}_{\vex{R},\alpha,\eta,s}c_{\vex{q},a',\eta',s'}|G\rangle=&0.
		\end{align}
	\end{subequations}

	In Ref.~\cite{Song2022}, the coupling constants are estimated as follows: $U=57.95$ meV, $W_{1,2}=44.03$ meV, $W_{3,4}=50.2$ meV, and $J=16.38$ meV. Since $U_2\ll U$, we ignore the $U_2$ term completely. We also note that the values of interaction terms are extracted with $w_0/w_1=0.8$, $w_1=110$ meV, and $\theta=1.05^{\circ}$ in the BM model \cite{Song2022}. 
	
	Now, we discuss the roles of the interaction terms. First, the $\hat{H}_U$ term denotes the onsite density-density interaction among $f$ fermions. We ignore the nearest-neighbor $U_2$ term as $U_2/U\approx 0.04\ll 1$ \cite{Song2022}. The $\hat{H}_V$ term describes the Coulomb interaction between $c$ fermions. The Hartree approximation of $H_V$ gives a correction to $\mu_c$ \cite{Song2022,Hu2023symmetric}, and the $c$-fermion band velocity may be slightly renormalized by the $\hat{H}_V$ term. However, the interaction effect of $\hat{H}_V$ is unlikely to induce a phase transition in $c$ fermions for the parameter regime of interest. The $\hat{H}_W$ term incorporates the density-density interaction between $c$ and $f$ fermions. 
	When the number of $f$ fermions per site is frozen to $N_f$, the density-density interaction is reduced to a bilinear term in $c$ fermions, $\hat{H}_W\rightarrow\Omega_0(N_f-4)\sum_{\vex{R},a}W_a:\hat{\rho}_{c,a}(\vex{R}):$. Since $W_1=W_2<W_3=W_4$, $\hat{H}_W$ with fixed $N_f$ per site can be viewed as a chemical potential term (which is absorbed in $\mu_c$) and a dynamical mass term $\propto (N_f-4)(W_3-W_1)$ in the $c$-fermion bands \cite{Zhou2023Kondo}. We will discuss the dynamical mass term later.
	Lastly, the $\hat{H}_J$ term is crucial for correlated insulating states in the Hartree-Fock calculations \cite{Song2022}, and $\hat{H}_J$ can be viewed as a $U(4)$ ferromagnetic Kondo coupling.

	Before moving forward, we comment on the values of $\mu_f$ and $\mu_c$. In the single-particle problem, one should consider $\mu_f=\mu_c$. However, in the presence of strong correlation, the values of $\mu_f$ and $\mu_c$ are renormalized due to the $\hat{H}_U$, $\hat{H}_V$, and $\hat{H}_W$ terms. In the Kondo-driven correlated metals, $\mu_f$ and $\mu_c$ are determined by self-consistent equations for given values of $\nu_f$ and $\nu_c$, the filling factors of $f$ and $c$ fermions, respectively. Thus, we ignore all the contributions to $\mu_f$ and $\mu_c$ but focus on the filling factors $\nu_f=4-N_f$ and $\nu_c$. As such, we ignore the contribution of $\hat{H}_V$ completely, and we only retain the dynamical mass contribution in $\hat{H}_W$, which will be discussed later.

\begin{figure}[t!]
		\includegraphics[width=0.4\textwidth]{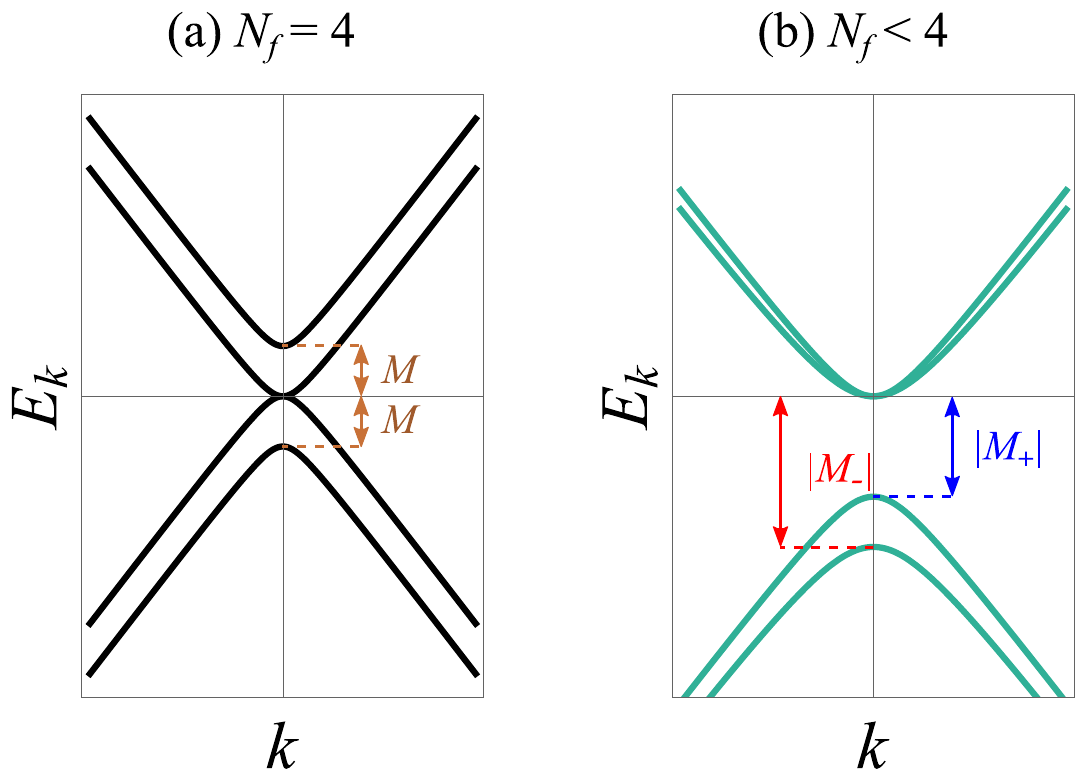}
		\caption{Band structures of $c$ fermions. (a) Single-particle bands described by Eq.~(\ref{Eq:H_0_c}). (b) $c$-fermion bands incorporating the dynamical mass contribution for $N_f< 4$ described by Eq.~(\ref{Eq:H_c_with_W}). $M_{\pm}=M\pm\bar{W}$.
		The $N_f>4$ case can be obtained by performing a particle-hole transformation.}
		\label{Fig:c_bands}
	\end{figure}

	With all the ingredients and approximations mentioned above, the THF model can be reduced to a generalized periodic Anderson model $\hat{H}_{\text{PAM}}=\hat{H}_{0,f}+\hat{H}_{0,c}+\hat{H}_{0,cf}+\hat{H}_U+\hat{H}_J$ [given by Eqs.~(\ref{Eq:H_0_f}), (\ref{Eq:H_0_c}), (\ref{Eq:H_U}), and (\ref{Eq:H_J_S})]. When $N_f$ is frozen and $N_f\neq 4$, we consider a dynamical ``mass'' term added to $\hat{H}_{0,c}$ such that
	\begin{align}\label{Eq:H_c_with_W}
		\hat{H}_{0,c}\rightarrow\sum_{\eta,s,a,a'}\sum_{\vex{q}}c^{\dagger}_{\vex{q},a,\eta,s}\left[h^{(\eta)}_{aa'}(\vex{q})+\mathcal{W}_{aa'}-\mu_c\delta_{aa'}\right]c_{\vex{q},a',\eta,s},
	\end{align}
where the dynamical mass term
	\begin{align}
		\hat{\mathcal{W}}=\left[\begin{array}{cc}
			\hat{0}_{2\times 2} & \hat{0}_{2\times 2}\\[2mm]
			\hat{0}_{2\times 2} & -\bar{W}\hat{1}_{2\times 2}
		\end{array}\right],
	\end{align}
	and $\bar{W}=-(N_f-4)(W_3-W_1)$. The definition of $\bar{W}$ such that $\bar{W}>0$ for $N_f<4$. We note that $\tr\hat{\mathcal{W}}\neq 0$, suggesting that $\hat{\mathcal{W}}$ contains a finite chemical potential shift. This choice of $\hat{\mathcal{W}}$ here is convenient for our calculations. Using a different choice of $\hat{\mathcal{W}}$ (along with a correspondingly modified $\mu_c$) gives the same results. In this work, we include the $\hat{\mathcal{W}}$ term and consider only the local-moment regime. We defer the complete treatment of $\hat{H}_W$ (incorporating valence fluctuation) to future work. The band structures are summarized in Fig.~\ref{Fig:c_bands}.
	
	\subsection{Single impurity model}
	
	At sufficiently high temperatures, we can treat the system as decoupled impurities of $f$ sites hybridized with the $c$ fermion band. In such a limit, we can reduce the problem to a surrogate single-impurity model, a proxy of the original lattice model, and apply the well-established poor man's scaling approach \cite{Anderson1970poorman,Hewson1997kondo,Coleman2015introduction}. 
	
	The surrogate impurity model is given by $\hat{H}_{\text{And}}=\hat{H}_{c,0}+\hat{H}_{\text{imp},f}+\hat{H}_{\text{imp},cf}$, where $\hat{H}_c$ is given by Eq.~(\ref{Eq:H_0_c}),
	\begin{align}
		\label{Eq:H_imp_f}\hat{H}_{\text{imp},f}=&-\mu_f\hat{n}_f+\frac{U}{2}\left(\hat{n}_f-4\right)^2,\\
		\label{Eq:H_imp_cf}\hat{H}_{\text{imp},cf}=&\frac{1}{\sqrt{N_s}}\sum_{\eta,s,\alpha,a}\!\sum_{\vex{q}}\left[V^{(\eta)}_{\alpha a}(\vex{q})f^{\dagger}_{\alpha,\eta,s}c_{\vex{q},a,\eta,s}\!+\!\text{H.c.}
		\right]\\
		\label{Eq:H_imp_J}\hat{H}_{\text{imp},J}=&-\frac{J}{\mathcal{N}_s}\sum_{\eta,\alpha,s_1,s_2}\sum_{\vex{q},\vex{q}'}e^{-\frac{\lambda^2\left(|\vex{q}|^2+|\vex{q}'|^2\right)}{2}}\\
		&\,\,\times\left[
		\begin{array}{c}
			:f^{\dagger}_{\alpha,\eta,s_1}f_{\alpha,\eta,s_2}::c^{\dagger}_{\vex{q}',\alpha+2,\eta,s_2}c_{\vex{q},\alpha+2,\eta,s_1}:\\[2mm]
			-f^{\dagger}_{\bar{\alpha},-\eta,s_1}f_{\alpha,\eta,s_2}c^{\dagger}_{\vex{q}',\alpha+2,\eta,s_2}c_{\vex{q},\bar{\alpha}+2,-\eta,s_1}
		\end{array}
		\right].
	\end{align}
	We consider the impurity at $\vex{R}=0$ without loss of generality and suppress all the position indices. 
	
	The impurity model in the decoupled limit (i.e., absence of $\hat{H}_{\text{imp},cf}$) can be solved exactly. The many-body energy level with $n$ fermions ($n=0,1,...,8$) is given by 
	\begin{align}
		E_n=-\mu_fn+\frac{U}{2}(n-4)^2.
	\end{align}
	The energy differences between nearby energy levels are as follows:
	\begin{align}
		E_{n-1}-E_n=&U\left(\frac{9}{2}-n\right)+\mu_f,\\
		E_{n+1}-E_n=&U\left(n-\frac{7}{2}\right)-\mu_f.
	\end{align}
	For a non-degenerate ground state with $N_f$ fermions, we derive
	\begin{align}
		N_f-\frac{9}{2}<\mu_f/U<N_f-\frac{7}{2}.
	\end{align}
	When $\mu_f/U=n-\frac{7}{2}$ with an integer $n=1-7$, $E_n$ and $E_{n+1}$ become the degenerate ground state levels. 
	
	In the presence of $\hat{H}_{\text{imp},cf}$, the charge fluctuations can renormalize the energy levels and result in a rich low-temperature phase diagram. In particular, one can derive the Kondo coupling via the Schrieffer-Wolf transformation provided that the ground state has exactly $N_f$ fermions ($N_f=1,2,...,7$) and $U\gg |\gamma|$. The Kondo impurity Hamiltonian is given by
	\begin{align}
		\nonumber\hat{H}_{\text{imp},K}=&\frac{J_K}{\mathcal{N}_s\gamma^2}\!\sum_{\substack{\alpha,\eta,a,s\\ \alpha',\eta',a',s'}}\sum_{\vex{q},\vex{q}'}\left[V^{(\eta')}_{\alpha'a'}(\vex{q}')\right]^*V^{(\eta)}_{\alpha a}(\vex{q})\\
		\label{Eq:H_K}&\!\times:\!\!f^{\dagger}_{\alpha,\eta,s}f_{\alpha',\eta',s'}\!::\!c^{\dagger}_{\vex{q}',a',\eta',s'}c_{\vex{q},a,\eta,s}:\hat{\mathcal{P}}_{N_f},
	\end{align}
	where $\hat{\mathcal{P}}_{N_f}$ is the projection operator onto the subspace with exactly $N_f$ localized $f$ fermions per site. The value of $J_K$ is determined by
	\begin{align}
		J_K=\gamma^2\left(\frac{1}{E_{N_f+1}-E_{N_f}}+\frac{1}{E_{N_f-1}-E_{N_f}}\right).
	\end{align}
	When $\mu_f=(N_f-4)U$, we obtain $J_K=4\gamma^2/U$ \cite{Chou2022kondo}. 
	
	In addition, a potential scattering term also arises during the Schrieffer-Wolf transformation,
	\begin{align}
		\nonumber\hat{H}_{\text{imp},\delta}=&\frac{\delta}{\mathcal{N}_s\gamma^2}\!\sum_{\alpha,\eta,a,a',s}\sum_{\vex{q},\vex{q}'}\sum_{\vex{R}}\left[V^{(\eta)}_{\alpha a'}(\vex{q}')\right]^*V^{(\eta)}_{\alpha a}(\vex{q})\\
		\label{Eq:H_P}&\!\times c^{\dagger}_{\vex{q}',a',\eta,s}c_{\vex{q},a,\eta,s}\hat{\mathcal{P}}_{N_f},
	\end{align}
	where
	\begin{align}
		\delta=\frac{\gamma^2}{2}\left(-\frac{1}{E_{N_f+1}-E_{N_f}}+\frac{1}{E_{N_f-1}-E_{N_f}}\right).
	\end{align}
	Since $\hat{H}_{\text{imp},\delta}$ contains only scatterings of $a=1,2$, it induces an imbalance of the local chemical potentials among $c$-fermion orbitals, similar to the role of $\hat{H}_W$ interaction locally. 

	In this work, we ignore the potential scattering terms and concentrate only on Kondo coupling and Hund's rule interaction. The potential impact due to $\hat{H}_{\text{imp},\delta}$ will be discussed in Sec.~\ref{Sec:Discussion}.
	Thus, the minimal surrogate impurity Kondo model is given by $\hat{H}_{\text{Kondo}}=\hat{H}_{0,c}+\hat{H}_{\text{imp},K}+\hat{H}_{\text{imp},J}$. We note that $\hat{H}_{\text{imp},K}$ and $\hat{H}_{\text{imp},J}$ have distinct matrix elements and act on different orbital sectors of $c$ fermions. Therefore, careful treatments for both $\hat{H}_{\text{imp},K}$ and $\hat{H}_{\text{imp},J}$ are required.
	
	\section{Poor man's scaling}\label{Sec:PMS}
	
	In this section, we apply the poor man's scaling \cite{Anderson1970poorman,Haldane1978,Hewson1997kondo,Coleman2015introduction,Cheng2017} to the single impurity Kondo model $\hat{H}_{\text{Kondo}}=\hat{H}_{0,c}+\hat{H}_{\text{imp},K}+\hat{H}_{\text{imp},J}$ and study the scaling flows of $J_K$ and $J$. The Kondo model here is derived when the number of localized $f$ fermions is frozen, i.e., a well-defined local moment is formed. In principle, one can determine local-moment and mixed-valence regimes by analyzing the impurity Anderson model. (See Appendix~\ref{Sec:App:Anderson_PMS} for a discussion on formalism.) In this work, as already mentioned, we concentrate only on the local-moment regime and investigate the Kondo coupling and Hund's rule interaction. 

	The poor man's scaling approach aims to provide a systematic understanding of the low-temperature physics with $k_BT\ll U/2$. The idea is to integrate out the high energy degrees of freedom and incorporate the renormalization effect in a progressive perturbative fashion. To begin, we consider the itinerant $c$ fermions with energy $E\in [-D,D]$, and then the energy intervals $[-D,-D+\delta D]$ and $[D-\delta D, D]$ are integrated out for an infinitesimal $\delta D$. Note that the largest $D$ should be less than $U/2$ for the validity of a well-defined local moment.
	Through the interactions between $c$ fermions and the local moment, the physical parameters are renormalized using the second-order perturbation theory. Unlike the standard renormalization group approach, the cutoff is not rescaled back to $D$ after each iteration in the poor man's scaling approach. Nevertheless, we can still derive flow equations for several physical quantities as functions of the running cutoff parameter (i.e., the half bandwidth $D$). Only the one-loop accuracy is incorporated in this work.

	\subsection{Scaling of $J_K$ term in the local moment regime}

	\begin{figure}[t]
		\includegraphics[width=0.45\textwidth]{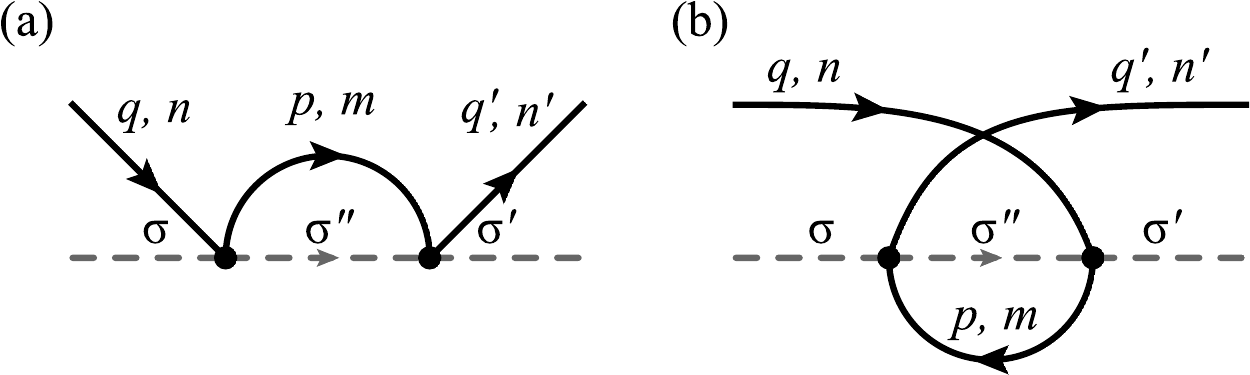}
		\caption{The one-loop diagram of poor man's scaling for Kondo coupling. (a) Particle contribution. (b) Hole contribution. $\sigma$, $\sigma'$, and $\sigma''$ indicate the local moment state; $q$, $q'$, and $p$ indicate the momentum of $c$ fermions; $n$, $n'$, and $m$ indicate the quantum number (spin, valley, and orbital) of $c$ fermions.}
		\label{Fig:T_matrix}
	\end{figure}	

	First, we focus on the scaling of Kondo coupling $J_K$. To derive the scaling equations, we employ the T-matrix formalism \cite{Hewson1997kondo,Coleman2015introduction}.
	We progressively integrate out the high energy $c$ fermions and reduce the half bandwidth from $D$ to $D-\delta D$ in each step. 
	This procedure produces corrections to $\hat{H}_{\text{Kondo}}=\hat{H}_{0,c}+\hat{H}_{\text{imp},K}$ through perturbing $\hat{H}_{\text{imp},K}$. The T-matrix Feynman diagrams at the one-loop level are illustrated in Fig.~\ref{Fig:T_matrix}. The detailed derivations can be found in Appendix.~\ref{Sec:App:TMatrix} We summarize the main results here. After reducing the half bandwidth from $D$ to $D-\delta D$, a correction to the low-energy Hamiltonian is generated:
	\begin{align}
		\delta \hat{H}=T^{(p)}+T^{(h)},
	\end{align}
	where 
	\begin{align}
		T^{(p)}\approx&\left[N_f J_K\delta D\sum_b\frac{\Delta_b(D+\mu_c)}{\pi\gamma^2 D}\right]\hat{H}_{\text{imp},K},\\
		T^{(h)}\approx&\left[(8-N_f) J_K\delta D\sum_b\frac{\Delta_b(-D+\mu_c)}{\pi\gamma^2 D}\right]\hat{H}_{\text{imp},K}.
	\end{align}
$T^{(p)}$ ($T^{(h)}$) is the contribution from the particle (hole) excitation in the $c$ fermion bands with the prefactor $N_f$ ($8-N_f$) indicating the number of the particle (hole) channels. The diagrams for the particle and hole contributions are illustrated in Fig.~\ref{Fig:T_matrix}. $\Delta_b$ describes the hybridization between $f$ fermion and the $b$th $c$-fermion band, given by 
\begin{align}
	\Delta_b(\mathcal{E})=\pi\rho_b(\mathcal{E})\Omega_0\left|\sum_aV^{(\eta)}_{\alpha a}(\vex{p})\Psi^{\eta}_{b,a}(\vex{p})\right|^2\Bigg|_{E_b(\vex{p})=\mathcal{E}},
\end{align}
where $\rho_b$ is the density of states associated with the $b$th band. The value of $\Delta_b(\mathcal{E})$ is independent of $\eta$ and $\alpha$ used in the calculations

When $\nu_c=0$ and $N_f=4$, the problem has a particle-hole symmetry, and the large-$D$ behavior is reminiscent of the pseudogap Kondo problem \cite{Ingersent1996,Gonzalez-Buxton1998,Fritz2004,Pixley2013,Cheng2017,Fritz2013physics}. Away from the special point, the particle and hole contributions are generically unequal, and potential scatterings [given by Eq.~(\ref{Eq:H_P})] arise. We discuss the potential scatterings in Sec.~\ref{Sec:Discussion} and Appendix~\ref{Sec:App:Imp}.

	Taking $\delta D\rightarrow 0$, one can obtain the one-loop poor man's scaling equation for $J_K$ given by
	\begin{align}
		\nonumber\frac{dJ_K}{dD}=&-N_fJ_K^2\left[\sum_b\frac{\Delta_b(D+\mu_c)}{\pi\gamma^2 D}\right]\\
		\label{Eq:dJ_KdD}&-(8-N_f)J_K^2\left[\sum_b\frac{\Delta_b(-D+\mu_c)}{\pi\gamma^2 D}\right].
	\end{align}
	We are interested in the flow of $J_K$ from an initial half bandwidth $D_i$ to $D_f$ ($D_i>D_f$). Equation~(\ref{Eq:dJ_KdD}) is reduced to the scaling flow for the $SU(8)$ Coqblin-Schrieffer model when $\Delta_b(\mathcal{E})$ is a constant \cite{Coleman2015introduction}. The minus sign in Eq.~(\ref{Eq:dJ_KdD}) indicates that $J_K$ flows up as $D$ decreases. The scaling equation here is strictly valid for $J_K(D)\Omega_0\rho_c(D)\ll 1$, where $\rho_c$ is the $c$-fermion density of states per spin per valley. Otherwise, higher-order corrections are required.
	
	To gain more insight about the flow of $J_K$, we integrate Eq.~(\ref{Eq:dJ_KdD}) and derive
	\begin{align}
		\nonumber J_K(D_f)=&\Bigg\{\frac{1}{J_K(D_i)}-N_f\int_{D_f}^{D_i}dD\left[\sum_b\frac{\Delta_b(D+\mu_c)}{\pi\gamma^2 D}\right]\\
		\label{Eq:JK_Df}&-(8-N_f)\int_{D_f}^{D_i}dD\left[\sum_b\frac{\Delta_b(-D+\mu_c)}{\pi\gamma^2 D}\right]\Bigg\}^{-1}.
	\end{align}
	For a sufficiently small $D_f$, $J_K(D_f)$ diverges, and the energy scale is associated with the Kondo temperature $D_K=k_BT_K$ within the one-loop accuracy. The Kondo energy scale can be obtained by solving the following equality:
	\begin{align}\label{Eq:Det_TK}
		\nonumber\frac{1}{J_K(D_i)}=&N_f\int_{D_K}^{D_i}dD\left[\sum_b\frac{\Delta_b(D+\mu_c)}{\pi\gamma^2 D}\right]\\
		&(8-N_f)\int_{D_K}^{D_i}dD\left[\sum_b\frac{\Delta_b(-D+\mu_c)}{\pi\gamma^2 D}\right].
	\end{align}
	Unlike the standard Kondo problems with constant density of states and constant couplings, the analytic expression of $D_K$ here is challenging due to the complicated expression of $\Delta_b$.
	Thus, we extract $D_K$ numerically for different filling factors and discuss the results in Sec.~\ref{Sec:Results}.

	\subsection{Scaling of $J$ term}

In addition to the Kondo coupling, it is also interesting to investigate the scaling of the Hund's rule interaction $J$ term. Since the $J$ term comes from the $c-f$ exchange Coulomb interaction, the matrix elements differ greatly from the Kondo coupling $J_K$ term. The derivations of T-matrix contributions are provided in Appendix~\ref{Sec:App:TMatrix}. We summarize the main results in the following.

At CNP (i.e., $N_f=4$, $\mu_c=0$), the scaling equation can be derived analytically and is given by
\begin{align}\label{Eq:dJdD_ph}
	\frac{dJ}{dD}=4J^2\left[\sum_{b}\frac{\bar{\Delta}_b(D)}{\pi D}\right],
\end{align}
where 
\begin{align}
	\bar{\Delta}_b(\mathcal{E})=\pi \rho_b(\mathcal{E})\Omega_0\left[e^{-\lambda^2|\vex{p}|^2}\left|\psi^{(\eta)}_{b,\alpha+2}(\vex{p})\right|^2\right]\bigg|_{E_b(\vex{p})=\mathcal{E}}
\end{align}
is the modified hybridization function of $b$th band.  
$\bar{\Delta}_b$ gives the same result for $\alpha=1,2$ and is independent of $\eta$. Equation~(\ref{Eq:dJdD_ph}) is reduced to the $SU(4)$ ferromagnetic Kondo model scaling when $\bar{\Delta}_b(\mathcal{E})$ is set to a constant. The plus sign in Eq.~(\ref{Eq:dJdD_ph}) indicates $J$ flows down as $D$ decreases.
For the general cases, we obtain an approximated scaling equation given by
\begin{align}
	\nonumber\frac{dJ}{dD}=&\frac{N_f}{2}J^2\sum_b\frac{\bar{\Delta}_b(D+\mu_c)}{\pi D}\\
	\label{Eq:dJdD_gen}&+\left(4-\frac{N_f}{2}\right)J^2\sum_b\frac{\bar{\Delta}_b(-D+\mu_c)}{\pi D}.
\end{align}
With $N_f=4$ and $\mu_c=0$, the expression in Eq.~(\ref{Eq:dJdD_gen}) is reduced to Eq.~(\ref{Eq:dJdD_ph}). Similar to the scaling flow for $J_K$, impurity potential scatterings arise away from CNP. The potential scattering term associated with $J$ is described by Eq.~(\ref{Eq:H_P'}) in Appendix~\ref{Sec:App:Imp}. Again, this leading order flow equation is valid when $J(D)\Omega_0\rho_c(D)\ll1$.

The scale-dependent $J$ can be obtained by integrating Eq.~(\ref{Eq:dJdD_gen}), and is given by
\begin{align}
	\nonumber J(D_f)=&\Bigg\{\frac{1}{J(D_i)}+\frac{N_f}{2}\int_{D_f}^{D_i}dD\left[\sum_b\frac{\bar{\Delta}_b(D+\mu_c)}{\pi D}\right]\\
	\label{Eq:J_Df}&+\frac{8-N_f}{2}\int_{D_f}^{D_i}dD\left[\sum_b\frac{\bar{\Delta}_b(-D+\mu_c)}{\pi D}\right]\Bigg\}^{-1},
\end{align}
where $D_i$ is the initial half bandwidth, $D_f$ is the final half bandwidth, and $D_i>D_f\ge0$. Note that $\bar{\Delta}_b$ has a different energy dependence than $\Delta_b$. For example, with $N_f=4$ and $\mu_c=0$, the integral $\int_{0}^{D_i} dy\bar{\Delta}_{3,4}(y)/y$ is finite. (The corresponding integral for $\Delta_{3,4}$ is always divergent.) As a result, we find that $J(D\rightarrow 0)$ remains finite when $\mu_c=0$ and $N_f=4$. Away from $N_f=4$ and $\mu_c=0$, $J(D\rightarrow 0)=0$ holds generally.

	\begin{figure}[t]
	\includegraphics[width=0.325\textwidth]{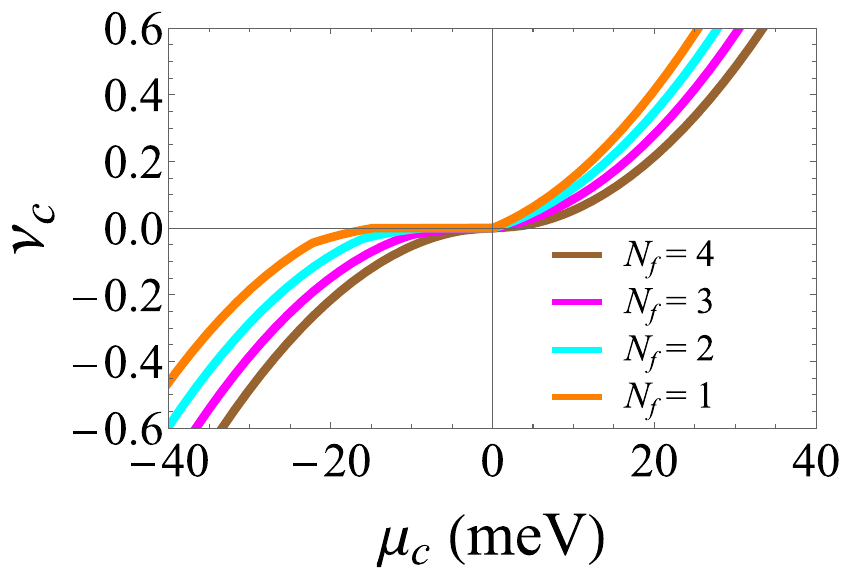}
	\caption{The filling factor $\nu_c$ as a function of $\mu_c$ with different $N_f$. The brown, magenta, cyan, orange lines represent $N_f=4,3,2,1$ respectively. The strong particle-hole asymmetry for $N_f\neq 4$ is a consequence of the dynamical mass due to the $W$ term. For $N_f=5,6,7$, one can simply use particle-hole transformation of $N_f=1,2,3$. }
	\label{Fig:nu_c}
\end{figure}	

	\section{Results}\label{Sec:Results}
	
	In this section, we present the numerical results based on the one-loop poor man's scaling equations derived in the previous section. In particular, we focus on the estimate of $T_K$ and the characteristic $J$ as functions of $N_f$ and $\nu_c$ (filling factor of $c$ fermions). The chemical potentials $\mu_f$ and $\mu_c$ are typically strongly modified in the self-consistent equations of the original Kondo lattice calculations \cite{Hewson1997kondo,Coleman2015introduction,Chou2022kondo}. Therefore, the dependence of filling factors provides more direct access to the MATBG experiments. 
	These results provide useful guidance for quantum many-body calculations for the Kondo lattice model for MATBG \cite{Chou2022kondo,Hu2023kondo,Hu2023symmetric}.

	Before we discuss the results, it is useful to emphasize that we treat $N_f$ and $\nu_c$ as theoretical tunable parameters. The precise values of $N_f$ and $\nu_c$ for a given total filling can be determined by free energy calculations for the Kondo lattice model for MATBG. However, our results provide a qualitative understanding of the filling dependence. 
	For $N_f=4$, we consider the particle-hole symmetric single-particle $c$-fermion bands without any spectral gap. For $N_f\neq 4$, a dynamical mass is generated in the local moment regime (where $N_f$ is fixed per site), and a spectral gap and particle-hole asymmetry develop. It is useful to convert $\mu_c$ to $\nu_c$ with the following formula:
	\begin{align}
		\nu_c(\mu_c)=4\Omega_0\int_{0}^{\mu_c}d\mathcal{E}\left[\sum_{b=1}^4\rho_b(\mathcal{E})\right],
	\end{align}
	where $\nu_c(\mu_c)$ is the $c$-fermion filling factor as a function of the chemical potential $\mu_c$, and the factor of 4 denotes the degeneracy factor of spin and valley. In the Kondo lattice problem, the total filling factor $\nu=(4-N_f)+\nu_c$ \cite{Chou2022kondo,Hu2023symmetric}. The poor man's scaling calculations here aim to resolving the $T_K$ and estimating the characteristic $J$ corresponding to the same filling in MATBG.
	The function $\nu_c(\mu_c)$ depends on the value of $N_f$ as shown in Fig.~\ref{Fig:nu_c} because of the dynamical mass in the $c$-fermion band. Again, the dynamical mass effect is due to the Kondo lattice model for MATBG, and we incorporate such an effect in the surrogate single-impurity model.

	We summarize all the parameters used in the following. $v_*=-4.303$ eV \AA, $M=3.697$ meV, $\gamma=-24.75$ meV, $v_*'=1.622$ eV \AA, $\lambda=0.3375a_M$, $a_M=134.24$ \AA, $U=57.95$ meV, $\bar{W}=W_3-W_1=6.17$ meV, $J_K=42.28$ meV, and $J=16.38$ meV. These parameters correspond to $w_0/w_1=0.8$, $w_1=110$ meV, and $\theta=1.05^{\circ}$ in the BM model \cite{Song2022}.

	In the rest of this section, we first discuss the general properties of the results near CNP. Then, we examine the filling-dependence of results using the parameters provided in Ref.~\cite{Song2022}.
	
		\begin{figure}[t]
		\includegraphics[width=0.325\textwidth]{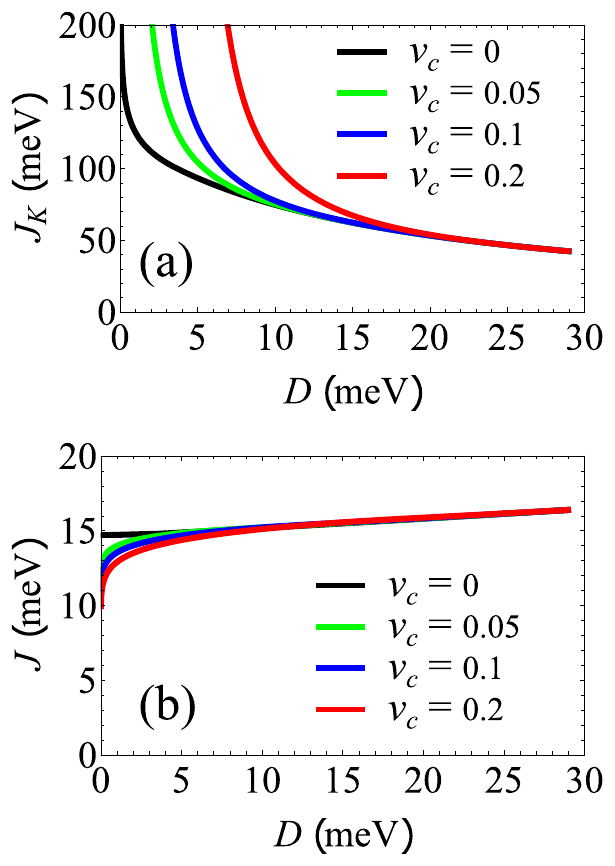}
		\caption{The poor man's scaling flows with $N_f=4$. (a) The Kondo coupling $J_K$ with the bare value $J_K(D_i)=42.28$meV and $D_i=U/2=28.9$meV. (b) The Hund's rule coupling $J$ with $J(D_i)=16.38$meV at $D_i=U/2=28.9$meV. Black lines indicate $\nu_c=0$; green lines indicate $\nu_c=0.05$; blue lines indicate $\nu_c=0.1$; red lines indicate $\nu_c=0.2$. The flows depend strongly on the value of $\nu_c$. $J_K\Omega_0\rho(D)<1$ and $J\Omega_0\rho(D)<1$ for all the results presented here.}
		\label{Fig:Flow}
	\end{figure}

	\subsection{Near charge neutrality point ($N_f=4$)}

	\begin{figure}[t]
	\includegraphics[width=0.45\textwidth]{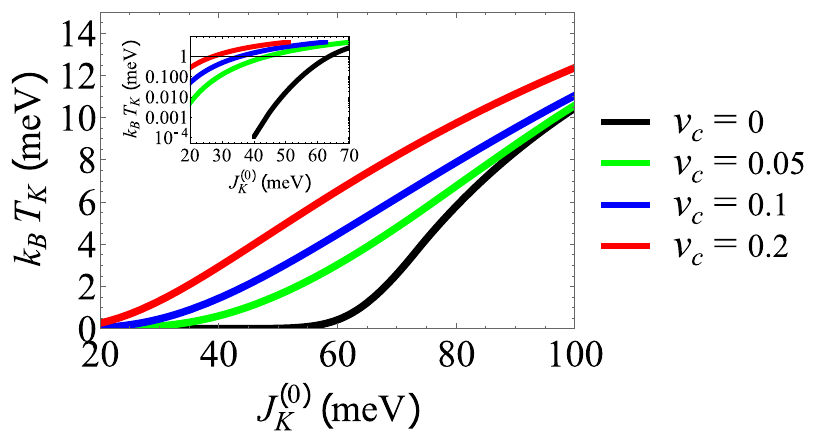}
	\caption{Kondo temperature as a function of $J_K^{(0)}$ with $N_f=4$. We run the scaling equation with $J_K^{(0)}\equiv J_K(D_i)$ and $D_i=U/2=28.9$meV. The Kondo temperature ($T_K$) is determined by Eq.~(\ref{Eq:Det_TK}) with $D_K=k_BT_K$. Black lines indicate $\nu_c=0$; green lines indicate $\nu_c=0.05$; blue lines indicate $\nu_c=0.1$; red lines indicate $\nu_c=0.2$. $T_K$ depends strongly on the values of $J_K^{(0)}$ and $\nu_c$.}
	\label{Fig:TK_vs_JK_Nf4}
\end{figure}

	In this subsection, we focus on the local moment regime with $N_f=4$ corresponding to $\nu_f=0$ in MATBG. In this case, the $c$-fermion bands are described by Eq.~(\ref{Eq:H_0_c}). We discuss the results with different values of $\nu_c>0$. Because of the particle-hole symmetry in the single-particle $c$-fermion bands, the results only depend on $|\nu_c|$.
	
	The poor man's scaling flows of $J_K$ and $J$ [given by Eqs.~(\ref{Eq:JK_Df}) and (\ref{Eq:J_Df})] with a few representative values of $\nu_c$ are plotted in Fig.~\ref{Fig:Flow}. We use the estimated values of $J_K$ and $J$ from Ref.~\cite{Song2022} and an initial half bandwidth $D_i=U/2=28.9$ meV. In Fig.~\ref{Fig:Flow}(a), the value of $J_K$ flows toward infinity for a sufficiently small $D$, indicating the onset of Kondo screening. Generically, the divergence manifests at a higher $D_K$ for a larger $|\nu_c|$, where $D_K$ is the energy scale indicating $J_K$ becomes divergent within the one-loop accuracy. The renormalization of $J$ shows the opposite trend. In Fig.~\ref{Fig:Flow}(b), the value of $J$ decreases as $D$ decreases. For $\nu_c\neq 0$, $J$ vanishes at $D=0$, suggesting $J$ is irrelevant under scaling flow for $\nu_c\neq 0$. For $\nu_c=0$, $J$ remains finite at $D=0$, and the suppression of $J$ may not be significant. This is due to the form of $\bar{\Delta}_b$ [Eq.~(\ref{Eq:Delta_bar_Nf_4})] such that the common low-energy logarithmic divergence in Eq.~(\ref{Eq:J_Df}) is absent for $\nu_c=0$. 
	
	Kondo temperature is often sensitive to the strength of Kondo coupling and the $c$-fermion density of states \cite{Hewson1997kondo,Coleman2015introduction}. To examine the Kondo temperature, we compute the energy scale $D_K$ at which $J_K(D_K)$ diverges for a wide range of initial values of $J_K^{(0)}\equiv J_K(D_i)$ with an initial half bandwidth $D_i=U/2=28.9$meV. In Fig.~\ref{Fig:TK_vs_JK_Nf4}, we show the Kondo temperature as a function of $J_K^{(0)}$ for different $\nu_c$. For $J_K^{(0)}>60$meV, we obtain $D_K=k_BT_K>0.5$meV for all cases. We find that the larger the $\nu_c$, the higher the value of $T_K$. For $40\text{meV}<J_K^{(0)}<60$meV, $T_K$ shows an exponential dependence on the value of $J_K^{(0)}$ as plotted in the inset of Fig.~\ref{Fig:TK_vs_JK_Nf4}. Particularly, using $J_K^{(0)}=42.28$meV, we obtain $D_K=k_BT_K\approx 4.14\times 10^{-4}$meV, which is comparable to the previous estimate in Ref.~\cite{Zhou2023Kondo}. We note that the above result does not rule out the possibility of a Kondo-driven semimetal \cite{Chou2022kondo} at charge neutrality of MATBG because the value $J_K^{(0)}$ is not precisely known, and the Kondo temperature estimate is very sensitive to $J_K$ as suggested in Fig.~\ref{Fig:TK_vs_JK_Nf4}. In addition, $T_K$ can be significantly enhanced by a small nonzero value of $\nu_c$. For example, using $J_K^{(0)}=42.28$meV and $\nu_c=0.05$, we obtain $k_BT_K\approx 0.8$meV, which is three orders of magnitude larger than the estimated $T_K$ at $\nu_c=0$.

	\begin{figure}[t]
		\includegraphics[width=0.325\textwidth]{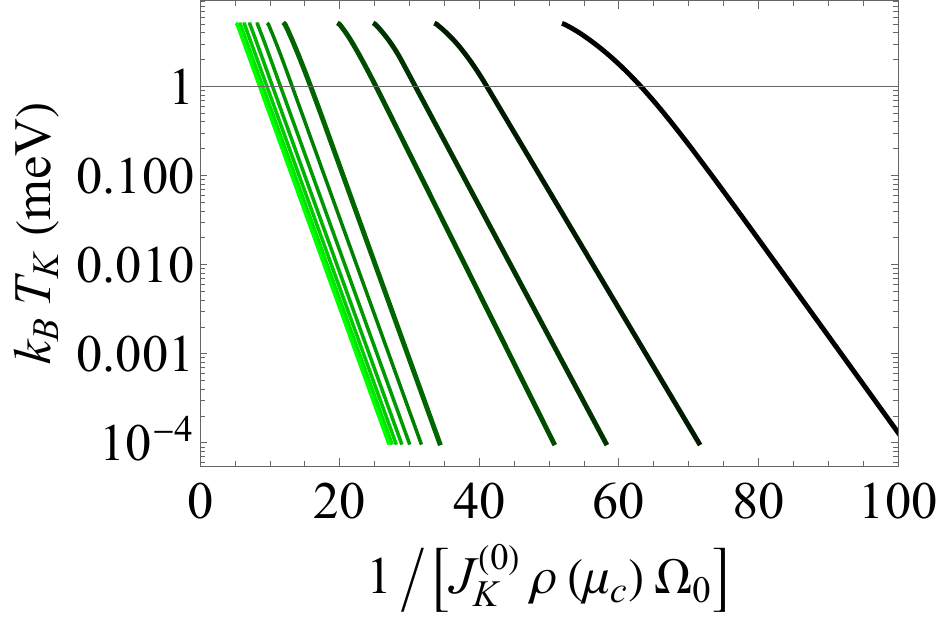}
		\caption{Exponential scaling of Kondo temperature for $N_f=4$. $T_K$ as a function of $1/\left[J_K^{(0)}\rho(\mu_c)\Omega_0\right]$ is plotted for different values of $\mu_c$. From right to left: $\mu_c=0$ (black line), $1$meV, $2$meV, 3meV, \dots, 10meV (green line). The results are obtained with $D_i=U/2=28.9$ meV.}
		\label{Fig:TK_fitting}
	\end{figure}	
	
	Another interesting question is whether the poor man's scaling produces the celebrated relation $T_K\propto\exp\left(-C/g\right)$ \cite{Hewson1997kondo,Coleman2015introduction}, where $C$ is some constant, and $g$ is the dimensionless Kondo coupling parameter. In Fig.~\ref{Fig:TK_fitting}, we plot $T_K$ as a function of $g\equiv 1/\left[J_K^{(0)}\rho(\mu_c)\Omega_0\right]$, where $\rho(\mu_c)$ is the $c$-fermion density of states per spin per valley and $\Omega_0$ is the moir\'e unit cell area. We find that the Kondo temperature generically follows the $T_K\propto\exp\left(-C/g\right)$ dependence for the parameter regime of interest. Moreover, the constant $C$ depends strongly on the value of $\mu_c$. We find that $C\approx1/4$ for $\mu_c=0$, and the value of $C$ continuously increases with $\mu_c$. For $\mu_c>M\approx 3.7$meV (corresponding to $\nu_c>0.0073$), $C$ saturates to $1/2$. Note that $\nu_c=0.05$ corresponds to $\mu_c\approx 9.65$meV. The above results again suggest strong filling factor dependence (or equivalently, $\mu_c$) of $T_K$.

		\begin{figure}[t]
		\includegraphics[width=0.45\textwidth]{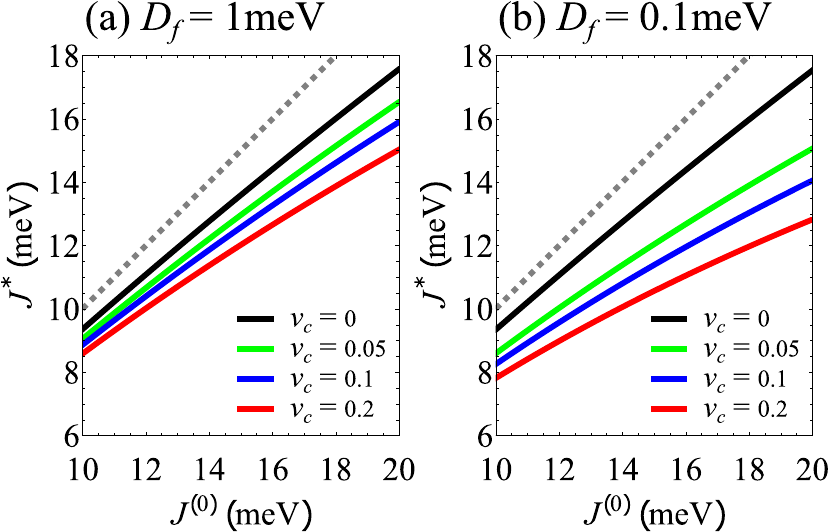}
		\caption{Characteristic Hund's rule coupling $J^*=J(D_f)$ with $N_f=4$. (a) $D_f=1$meV. (b) $D_f=0.1$meV. The results are obtained with $D_i=U/2=28.9$ meV. Black lines indicate $\nu_c=0$; green lines indicate $\nu_c=0.05$; blue lines indicate $\nu_c=0.1$; red lines indicate $\nu_c=0.2$; gray dashed lines indicate $J^*=J^{(0)}$.}
		\label{Fig:J_Df}
	\end{figure}

	Finally, we discuss Hund's rule coupling. In Fig.~\ref{Fig:J_Df}, we consider different values of $J^{(0)}=J(D_i)$ (with $D_i=U/2=28.9$meV) and plot the value of $J^*=J(D_f)$ at a much lower scale $D_f$ [$D_f=1$meV in Fig.~\ref{Fig:J_Df}(a) and $D_f=0.1$meV in Fig.~\ref{Fig:J_Df}(b)]. $J^*$ is always reduced from the bare value $J^{(0)}$, and the reduction is stronger for a larger $|\nu_c|$ or a smaller $D_f$. Based on the scaling flow [Eq.~(\ref{Eq:dJdD_gen})], $J(D)$ typically vanishes as $D\rightarrow 0$ (except for $\nu_c=0$). With $D_f\ge 0.1$meV, the value of $J^*$ is not substantially smaller than $J^{(0)}$ for the parameters we explored here. The weaker renormalization of $J$ (as compared to $J_K$) can be understood through the expression of $\bar{\Delta}_b$ Eq.~(\ref{Eq:Delta_bar_Nf_4}), which has a much weaker logarithmic infrared divergence at low energy as compared to $\Delta_b$.

	\subsection{Filling factor dependence}

	\begin{figure}[t]
	\includegraphics[width=0.325\textwidth]{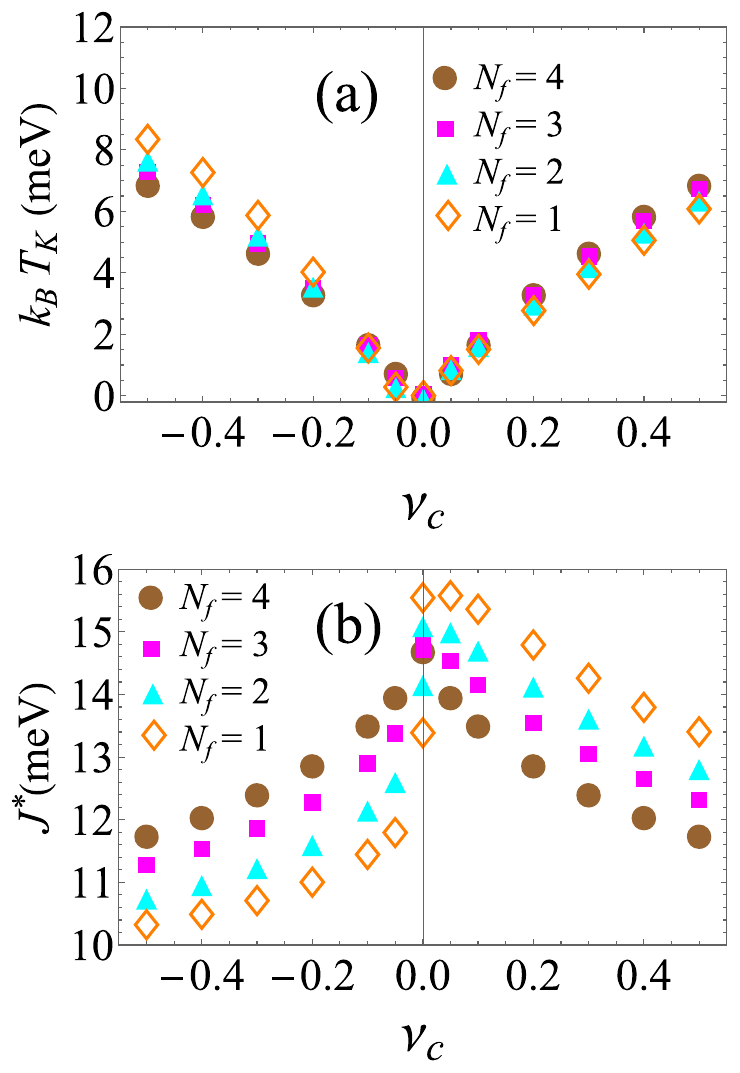}
	\caption{Kondo temperature ($T_K$) and characteristic Hund's rule coupling ($J^*$) as functions of $N_f$ and $\nu_c$ incorporating the dynamical mass $\bar{W}$ term. (a) Kondo temperature as a function of $\nu_c$. (b) Characteristic Hund's rule coupling as a function of $\nu_c$. $J^*$ is set by $J(D_f=1\text{meV})$. The results are obtained using $J_K(D_i)=42.28$meV, $J(D_i)=16.38$meV, and $D_i=U/2=28.9$meV. For $N_f\neq 4$, we incorporate the dynamical mass term $\bar{W}$ in the $c$-fermion bands. Brown solid dots indicate $N_f=4$; magenta solid square indicate $N_f=3$; cyan solid triangles indicate $N_f=2$; orange opened diamonds indicate $N_f=1$.
	}
	\label{Fig:TK_J_nu_Nf}
\end{figure}	

		\begin{figure}[t]
	\includegraphics[width=0.325\textwidth]{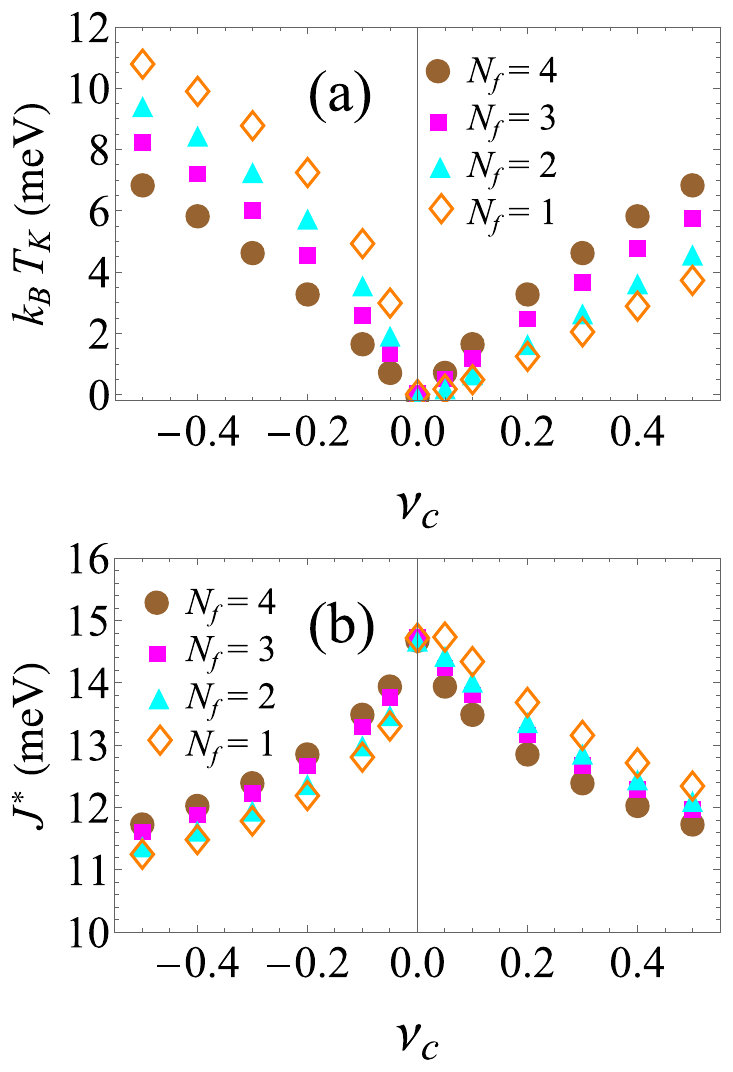}
	\caption{Kondo temperature ($T_K$) and characteristic Hund's rule coupling ($J^*$) as functions of $N_f$ and $\nu_c$ without the dynamical mass $\bar{W}$ term. (a) $T_K$ as a function of $N_f$ and $\nu_c$. (b) $J^*$ as a function of $N_f$ and $\nu_c$. $J^*$ is set by $J(D_f=1\text{meV})$. The results are obtained using $J_K(D_i)=42.28$meV, $J(D_i)=16.38$meV, and $D_i=U/2=28.9$meV. We consider the bare $c$-fermion bands without $\bar{W}$ dynamical mass. Brown solid dots indicate $N_f=4$; magenta solid square indicate $N_f=3$; cyan solid triangles indicate $N_f=2$; orange opened diamonds indicate $N_f=1$.
	}
	\label{Fig:TK_J_nu_Nf_bare}
\end{figure}	

	When $N_f\neq4$, the density-density interaction [the $H_W$ term given by Eq.~(\ref{Eq:H_W})] between $c$ and $f$ fermions can produce a strong renormalization of the $c$-fermion bands. In the local moment regime (i.e., $N_f$ is fixed), the $c$-fermion dispersion acquires a dynamical mass term as described by Eq.~(\ref{Eq:H_c_with_W}). Consequently, a spectral gap manifests and the dispersion is particle-hole asymmetric. We incorporate the modified $c$-fermion bands into the poor man's scaling calculations for $N_f\neq 4$. In this section, we focus on $J_K(D_i)=42.28$meV, $J(D_i)=16.38$meV, and $D_i=U/2=28.9$meV. Formally, each interaction term is renormalized as the half bandwidth is depleted, and the dynamical mass should vary in the process. For simplicity, we use $W_3-W_1=6.17$meV and ignore the renormalization of $\bar{W}$ for all the calculations.
	
	In Fig.~\ref{Fig:TK_J_nu_Nf}, Kondo temperature, $T_K$, and the characteristic Hund's rule coupling, $J^*=J(1\text{meV})$, are computed as functions of $N_f$ and $\nu_c$. For $N_f=4$, the results are symmetric under $\nu_c\rightarrow-\nu_c$ operation because of the particle-hole symmetry. For $N_f\neq 4$, the results show significant asymmetry in $\nu_c$, which is enhanced for a larger $|4-N_f|$. Specifically, $T_K$ with $N_f<4$ is typically larger (smaller) than $N_f=4$ for $\nu_c<0$ ($\nu_c>0$), and $J^*$ with $N_f<4$ is smaller (larger) than $N_f=4$ for $\nu_c<0$ ($\nu_c>0$). The results suggest that the renormalization is typically stronger (weaker) for $\nu_c<0$ ($\nu_c>0$) with $N_f<4$. Meanwhile, for $-0.1\le\nu_c<0$, $T_K$ with $N_f<4$ is slightly smaller than $N_f=4$, suggesting a weaker renormalization of $J_K$ in this small parameter regime.

	There are two sources that cause the asymmetry in $\nu_c$. First, the one-loop poor man's scaling contributions can be classified by the particle and hole channels, which are proportional to the available numbers of particles ($N_f$) and holes ($8-N_f$), respectively. When $\rho_c(D+\mu_c)>\rho_c(-D+\mu_c)$ ($\rho_c(D+\mu_c)<\rho_c(-D+\mu_c)$), the single-particle contribution is generally stronger (weaker) than the single-hole contribution. The overall contribution is $N_f$ times the single-particle contribution plus $8-N_f$ times the single-hole contribution. Therefore, for $N_f<4$, the renormalization of $\nu_c<0$ is typically stronger than for $\nu_c>0$. Second, the dynamical mass $\bar{W}$ generates asymmetric $c$-fermion bands and a spectral gap. For $\nu_c<0$ ($\nu_c>0$), the contribution from the particle (hole) channel is strongly suppressed by the spectral gap. Thus, the presence of spectral gap enhances (suppresses) the renormalization for $\nu_c<0$ ($\nu_c>0$) as long as $N_f<4$. The structure of the one-loop contribution and the presence of the spectral gap explain the overall trend of the results: The renormalization of $\nu_c<0$ is mostly stronger than $\nu_c>0$ for $N_f<4$, and the asymmetry is the strongest for $N_f=1$. While the above ideas can explain most of the results, $T_K$ for $N_f<4$ in the regime $-0.1\le\nu_c<0$ cannot be accounted for by the above arguments. The counterintuitive results in this regime are due to the asymmetric $c$-fermion bands and the structure of $\Delta_b$. On the other hand, this subtle feature is absent in $J^*$.

 	Finally, it is interesting to examine if the dynamical mass $\bar{W}$ term is essential for the estimated $T_K$ and $J^*$. In Fig.~\ref{Fig:TK_J_nu_Nf_bare}, we compute $T_K$ and $J^*(1\text{meV})$ as functions of $N_f$ and $\nu_c$ without the $\bar{W}$ term. We find the similar asymmetry in $\nu_c$, suggesting stronger renormalization for $\nu_c<0$ and weaker renormalization for $\nu_c>0$ with $N_f<4$. The results can be explained by the one-loop poor man's scaling contribution, and the $c$-fermion bands are completely particle-hole symmetric. It is worth noting that the asymmetry in $T_K$ ($J^*$) is slightly stronger (weaker) without the $\bar{W}$ term.

	\section{Discussion}\label{Sec:Discussion}
	
	Using the one-loop poor man's scaling approach and inspired by the THF model \cite{Song2022}, we derive analytical scale-dependent intrinsic Kondo ($J_K$) and Hund's rule ($J$) interactions in a surrogate single-impurity model for MATBG. We find that Kondo temperature ($T_K$) varies sensitively with the value of $J_K$ and the filling factor for the parameter regime associated with MATBG experiments. We also provide a detailed account of the characteristic value of $J$ as a function of filling factor. Note that the poor man's scaling calculations with given $N_f$ and $\nu_c$ can be interpreted as the filling $\nu=(4-N_f)+\nu_c$ in MATBG, and the dynamical mass due to $\hat{H}_W$ term is also incorporated for $N_f\neq 4$.
	Our results systematically estimate the renormalized interactions and $T_K$, which is essential for many-body calculations in MATBG. Moreover, we anticipate that Kondo-driven metals likely exist for a wide range of fractional fillings as $k_BT_K>1$mK generally, and $J^*$ is always suppressed from the bare value. Lastly, the strong filling-factor dependence of $T_K$ and $J^*$ suggests that correlation can be controlled by an external gate (i.e., changing filling). This effect is primarily due to the density of states of the $c$-fermion bands and is similar to the recent studies of single vacancy in graphene \cite{May2018} and single magnetic impurity coupled to twisted bilayer graphene \cite{Shankar2023}, where nontrivial density of states arises.
	
	A particularly interesting question is whether Kondo-driven semimetal \cite{Chou2022kondo} exists near the CNP in MATBG. In Fig.~\ref{Fig:TK_vs_JK_Nf4}, we show that $T_K$ with $N_f=4$ and $\nu_c=0$ varies strongly for 40mK$<J_K<$60mK. Using the estimated parameters in Fig.~\cite{Song2022}, we find $J_K\approx 42.28$meV, corresponding to $k_BT_K\approx4.14\times10^{-4}$meV, comparable to a previous estimate in Ref.~\cite{Zhou2023Kondo}. With $J_K\approx 60$meV, we obtain $k_BT_K\approx 0.5$meV, significantly larger than the estimate using $J_K\approx42.28$meV. The results suggest an extreme $J_K$ dependence in the value of $T_K$. Moreover, small doping (e.g., $\nu_c=0.05$) can significantly boost $T_K$ as shown in Fig.~\ref{Fig:TK_vs_JK_Nf4}. Note that $T_K$ corresponds to the energy scale of Kondo screening, and the Kondo-driven phases in MATBG happen at a lower energy, corresponding to a coherent off-diagonal long-range order. In addition, the correlated insulating states may preempt the Kondo-driven phases, which $ J$ governs. Assuming that $T_c$ is the transition temperature of the correlated insulator, the condition of realizing a Kondo-driven phase is $T_K> T_c$. Since $T_K$ is very sensitive to $J_K$ and $\nu_c$, and $J$ is always suppressed at low energies, the existence of a Kondo-driven semimetal near the CNP cannot be ruled out. The strong parameter dependence of $T_K$ also suggests that $T_K$ can vary significantly over different samples. The correlated insulating states dominate for $T_c>T_K$, while Kondo-driven topological semimetal can occur for $T_K> T_c$. Our analysis near CNP suggests that the gapped versus semimetallic behaviors in different MATBG experiments may depend on the variation of $J_K$ (and $J$). If this idea is correct, then Kondo physics plays an extremely important qualitative role in the physics of MATBG.
	
	Now, we discuss the implication of nonzero integer fillings in MATBG. First, the $c$-fermion dispersion is modified by a dynamical mass term originating from $\hat{H}_W$. The modified $c$-fermion band does not preserve particle-hole symmetry. Second, particle and hole contributions to the scaling are different. Generally, the contribution from the hole channel dominates for $N_f<4$. The overall trend for $N_f<4$ is that the $\nu_c<0$ regime has a stronger renormalization than the $\nu_c>0$ regime (except for $-0.1\le\nu_c\le 0$) as we show in Fig~\ref{Fig:TK_J_nu_Nf}. Assuming $\mu_c=0$ (corresponding to $\nu_c=0^+$) and $J_K=42.28$meV, we estimate the $T_K$ value corresponding to $\nu=-1,-2,-3$ in MATBG (corresponding to $N_f=3,2,1$). We find that $k_BT_K\approx 9\times 10^{-4}$meV for $\nu=-1$, $k_BT_K\approx 10^{-3}$meV for $\nu=-2$, and $k_BT_K\approx 1.99\times 10^{-6}$meV for $\nu=-3$. The values of $T_K$ with $\nu_c=0^-$ are smaller than the above estimates. Again, a nonzero $|\nu_c|$ and a larger $J_K$ \cite{Hu2023kondo} can boost $T_K$ significantly. Finally, potential scattering terms arise generically due to the imbalance between particle and hole contributions. The occurrence of potential scattering may hinder our quantitative estimates for nonzero integer fillings.

	In this work, we focus only on the local-moment regime where charge fluctuation is completely suppressed. In principle, one can study the poor man's scaling of the Anderson model and analyze various regimes, including local-moment and mixed valence regimes. Moreover, studying the scaling of the Anderson model can provide a more quantitative estimate of interaction parameters. In Appendix~\ref{Sec:App:Anderson_PMS}, we derive the poor man's scaling flows for each many-body level. We discuss a few generalities in the following. Similar to the $SU(2)$ Anderson model, the scaling flows tend to suppress $E_{N_f\pm1}-E_{N_f}$, where $N_f$ is the number of $f$ fermions on the impurity site. In Ref.~\cite{Song2022}, the THF model is valid with a half bandwidth of around 80 meV, larger than the $D_i=28.9$ meV used in the scaling calculations. Thus, we generally expect that $E_{N_f\pm1}-E_{N_f}$ is reduced from the bare value at $D_i$, and $J_K$ is enhanced concomitantly. The possibility that the actual $T_k$ may be larger because of variations in the effective parameters, therefore, cannot be ruled out and should be further investigated.
	
	An outstanding issue of our analysis is the generation of impurity potential scattering. At CNP, the potential scattering may arise from Eq.~(\ref{Eq:H_P}), and the value of $\delta$ remains constant during the poor man's scaling procedure, suggesting a line of fixed points analogous to the asymmetric $SU(2)$ Anderson model \cite{KrishnaMurthy1980}. Away from CNP, potential scatterings generally arise during the poor man's scaling calculations.
	In Appendix~\ref{Sec:App:Imp}, we derive the one-loop scaling flows in Eq.~(\ref{Eq:ddelta_dD}) for $\delta$ and $\delta'$ potential scattering terms. The $\delta$ term describes the scattering of $c$ fermions with orbital index $a=1,2$, while the $\delta'$ term describes the scattering of $c$ fermions with orbital index $a=3,4$. Since $J_K>J$ is generally true, the overall effect of these potential scatterings induces a mass term similar to the dynamical mass term from the interaction. For $N_f<4$, we find that the effect of potential scattering tends to enhance the dynamical mass term from interaction, i.e., the spectral gap of $c$ fermions gets larger as $D$ is decreased. We note that the one-loop equations cannot fully describe the $\delta$ term as $J_K$ diverges for a sufficiently small $D_f$. Going beyond one-loop results is necessary to understand the role of potential scattering.

	We conclude by discussing several possible generalizations of this work. The current analysis is based on the perturbative one-loop poor man's scaling approach, which is strictly valid for $J_K\rho_c(D)\Omega_0\ll 1$. It might be interesting to study higher-order corrections and perhaps the nonperturbative numerical renormalization group for a better quantitative understanding of $T_K$ and $\delta$. Moreover, the analysis based on the surrogate single-impurity model may not capture all the important features in the Kondo lattice for MATBG, which is our primary interest. To this end, a systematic analysis using lattice methods such as the dynamical mean field theory may be necessary. The analysis of this work can be generalized to magic-angle twisted trilayer graphene, where an analytical THF formulation has been obtained recently \cite{Yu2023magic}. In the trilayer graphene system, there is another type of itinerant fermions, the $d$ fermions, which hybridize with the localized $f$ fermions through the displacement field. We anticipate that the scaling flows are much richer in the trilayer graphene case as the problem is akin to a two-channel Kondo problem. 
	
		\begin{acknowledgments}
		\textit{Acknowledgments.---} Y.-Z.C. thanks Jed Pixley for useful discussions.
		This work is supported by the Laboratory for Physical Sciences. 
	\end{acknowledgments}

	\appendix
	
	\section{Delocalized $c$ fermion bands}
	
	In this section, we discuss the energy bands and wavefunction of the itinerant $c$ fermions. First, we discuss the bare $c$ fermion dispersion. Then, we incorporate the effect of a dynamical mass due to $\hat{H}_W$.
	
	\subsection{Single-particle bands}\label{Sec:App:c_band_bare}
	
	The bands are determined by the $\vex{k}\cdot\vex{p}$ Hamiltonian,
	\begin{align}
		\nonumber&\hat{h}^{(\eta)}(\vex{q})=\left[\!\begin{array}{cc}
			\hat{0}_{2\times 2} & v_*\!\left(\eta q_x\hat{\sigma}_0+iq_y\hat{\sigma}_z\right) \\[2mm]
			v_*\!\left(\eta q_x\hat{\sigma}_0-iq_y\hat{\sigma}_z\right) & M\hat{\sigma}_x		
		\end{array}\!\right]\\[2mm]
	=&\left[\begin{array}{cccc}
			0 & 0 & v_*|\vex{q}|e^{i\theta_{\eta,\vex{q}}} &0 \\
			0 & 0 &0 & v_*|\vex{q}|e^{-i\theta_{\eta,\vex{q}}} \\
			v_*|\vex{q}|e^{-i\theta_{\eta,\vex{q}}}& 0 &0 &M \\
			0 & v_*|\vex{q}|e^{i\theta_{\eta,\vex{q}}} &M &0 
		\end{array}
		\right],
	\end{align}
	where the angle $\theta_{\eta,\vex{q}}$ is defined by  $\eta k_x+ik_y\equiv |\vex{q}|e^{i\theta_{\eta,\vex{q}}}$. $\hat{h}^{(\eta)}$ can be diagonalized analytically using Mathematica. The expression of the energy bands are given by
	\begin{subequations}\label{Eq:App:E_c}
		\begin{align}
			E_1(\vex{q})=&\frac{1}{2}\left(-M-\sqrt{M^2+4v_*^2|\vex{q}|^2}\right),\\
			E_2(\vex{q})=&\frac{1}{2}\left(M-\sqrt{M^2+4v_*^2|\vex{q}|^2}\right),\\
			E_3(\vex{q})=&\frac{1}{2}\left(-M+\sqrt{M^2+4v_*^2|\vex{q}|^2}\right),\\
			E_4(\vex{q})=&\frac{1}{2}\left(M+\sqrt{M^2+4v_*^2|\vex{q}|^2}\right).
		\end{align}
	\end{subequations}
The energy bands are independent of valley, so the valley index is ignored. The corresponding wavefunctions are given by
\begin{subequations}\label{Eq:App:Psi_c}
	\begin{align}
		\Psi_1^{(\eta)}(\vex{q})=&\frac{1}{\sqrt{2}}\left[u_1(\vex{q}), -u_1^*(\vex{q}), -v_1(\vex{q}),\, v_1^*(\vex{q})
		\right]^T,\\
		\Psi_2^{(\eta)}(\vex{q})=&\frac{1}{\sqrt{2}}\left[-u_2(\vex{q}), -u_2^*(\vex{q}),\, v_2(\vex{q}),\, v_2^*(\vex{q})
		\right]^T,\\
		\Psi_3^{(\eta)}(\vex{q})=&\frac{1}{\sqrt{2}}\left[-u_3(\vex{q}),\, u_3^*(\vex{q}), -v_3(\vex{q}),\, v_3^*(\vex{q})
		\right]^T,\\
		\Psi_4^{(\eta)}(\vex{q})=&\frac{1}{\sqrt{2}}\left[u_4(\vex{q}),\, u_4^*(\vex{q}),\, v_4(\vex{q}),\, v_4^*(\vex{q})
		\right]^T,
	\end{align}
\end{subequations}
where
\begin{subequations}\label{Eq:App:u_v}
	\begin{align}
		u_1(\vex{q})=&\frac{\frac{1}{2}\left(-M\!+\!\sqrt{M^2\!+\!4v_*^2|\vex{q}|^2}\right)e^{i\theta_{\eta,\vex{q}}}}{\sqrt{\frac{1}{4}\!\left(\!-M\!+\!\sqrt{M^2\!+\!4v_*^2|\vex{q}|^2}\right)^2\!\!+\!v_*^2|\vex{q}|^2}},\\
		v_1(\vex{q})=&\frac{v_*|\vex{q}|}{\sqrt{\frac{1}{4}\!\left(\!-M\!+\!\sqrt{M^2\!+\!4v_*^2|\vex{q}|^2}\right)^2\!\!+\!v_*^2|\vex{q}|^2}},\\
		u_2(\vex{q})=&\frac{\frac{1}{2}\left(M\!+\!\sqrt{M^2\!+\!4v_*^2|\vex{q}|^2}\right)e^{i\theta_{\eta,\vex{q}}}}{\sqrt{\frac{1}{4}\!\left(\!M\!+\!\sqrt{M^2\!+\!4v_*^2|\vex{q}|^2}\right)^2\!\!+\!v_*^2|\vex{q}|^2}},\\
		v_2(\vex{q})=&\frac{v_*|\vex{q}|}{\sqrt{\frac{1}{4}\!\left(\!M\!+\!\sqrt{M^2\!+\!4v_*^2|\vex{q}|^2}\right)^2\!\!+\!v_*^2|\vex{q}|^2}},\\
		u_3(\vex{q})=&u_2(\vex{q}),\,\,\,v_3(\vex{q})=v_2(\vex{q}),\\
		u_4(\vex{q})=&u_1(\vex{q}),\,\,\,v_4(\vex{q})=v_1(\vex{q}).
	\end{align}
\end{subequations}
Additionally, we obtain
\begin{subequations}\label{Eq:App:VPsi}
	\begin{align}
		\hat{V}^{(\eta)}\Psi_1^{(\eta)}=&\frac{1}{\sqrt{2}}\left|u_1(\vex{q})\right|\left[\begin{array}{r}
			\gamma e^{i\theta_{\eta,\vex{q}}}-v'_*|\vex{q}|e^{-i2\theta_{\eta,\vex{q}}}\\[1mm]
			-\gamma e^{i\theta_{\eta,\vex{q}}}+v'_*|\vex{q}|e^{-i2\theta_{\eta,\vex{q}}}
		\end{array}\right],\\
		\hat{V}^{(\eta)}\Psi_2^{(\eta)}=&\frac{1}{\sqrt{2}}\left|u_2(\vex{q})\right|\left[\begin{array}{r}
			-\gamma e^{i\theta_{\eta,\vex{q}}}-v'_*|\vex{q}|e^{-i2\theta_{\eta,\vex{q}}}\\[1mm]
			-\gamma e^{i\theta_{\eta,\vex{q}}}-v'_*|\vex{q}|e^{-i2\theta_{\eta,\vex{q}}}
		\end{array}\right],\\
		\hat{V}^{(\eta)}\Psi_3^{(\eta)}=&\frac{1}{\sqrt{2}}\left|u_3(\vex{q})\right|\left[\begin{array}{r}
			-\gamma e^{i\theta_{\eta,\vex{q}}}+v'_*|\vex{q}|e^{-i2\theta_{\eta,\vex{q}}}\\[1mm]
			\gamma e^{i\theta_{\eta,\vex{q}}}-v'_*|\vex{q}|e^{-i2\theta_{\eta,\vex{q}}}
		\end{array}\right],\\
		\hat{V}^{(\eta)}\Psi_4^{(\eta)}=&\frac{1}{\sqrt{2}}	\left|u_4(\vex{q})\right|\left[\begin{array}{r}
			\,\gamma e^{i\theta_{\eta,\vex{q}}}+v'_*|\vex{q}|e^{-i2\theta_{\eta,\vex{q}}}\\[1mm]
			\,\gamma e^{i\theta_{\eta,\vex{q}}}+v'_*|\vex{q}|e^{-i2\theta_{\eta,\vex{q}}}
		\end{array}\right],
	\end{align}
\end{subequations}

The density of states per spin per valley for each band can be derived as follows:
\begin{subequations}\label{Eq:App:DOS_c}
		\begin{align}
		\rho_1(\mathcal{E})=&\Theta(-\mathcal{E}-M)\frac{-2\mathcal{E}-M}{4\pi v_*^2},\\
		\rho_2(\mathcal{E})=&\Theta(-\mathcal{E})\frac{-2\mathcal{E}+M}{4\pi v_*^2},\\
		\rho_3(\mathcal{E})=&\Theta(\mathcal{E})\frac{2\mathcal{E}+M}{4\pi v_*^2},\\
		\rho_4(\mathcal{E})=&\Theta(\mathcal{E}-M)\frac{2\mathcal{E}-M}{4\pi v_*^2},		
	\end{align}
\end{subequations}
where $\Theta(x)$ is the Heaviside function.

The hybridization between $f$ fermion and the $b$th $c$ fermion band, is described by $\Delta_b$ as follows:
\begin{subequations}\label{Eq:Delta_Nf_4}
	\begin{align}
		\nonumber\Delta_1(\mathcal{E})=&\Theta(-\mathcal{E}\!-\!M)\frac{-\mathcal{E}-M}{8v_*^2}\Omega_0e^{-\frac{\lambda^2}{v_*^2}(\mathcal{E}^2+\mathcal{E}M)}\\
		&\times\left[\gamma^2\!+\!\frac{v_*'^2}{v_*^2}(\mathcal{E}^2+\mathcal{E}M)\right]\!,\\
		\nonumber\Delta_2(\mathcal{E})=&\Theta(-\mathcal{E})\frac{-\mathcal{E}+M}{8v_*^2}\Omega_0e^{-\frac{\lambda^2}{v_*^2}(\mathcal{E}^2-\mathcal{E}M)}\\
		&\times\left[\gamma^2+\frac{v_*'^2}{v_*^2}(\mathcal{E}^2-\mathcal{E}M)\right],\\
		\nonumber\Delta_3(\mathcal{E})=&\Theta(\mathcal{E})\frac{\mathcal{E}+M}{8v_*^2}\Omega_0e^{-\frac{\lambda^2}{v_*^2}(\mathcal{E}^2+\mathcal{E}M)}\\
		&\times\left[\gamma^2+\frac{v_*'^2}{v_*^2}(\mathcal{E}^2+\mathcal{E}M)\right],\\		\nonumber\Delta_4(\mathcal{E})=&\Theta(\mathcal{E}-M)\frac{\mathcal{E}-M}{8v_*^2}\Omega_0e^{-\frac{\lambda^2}{v_*^2}(\mathcal{E}^2-\mathcal{E}M)}\\
		&\times\left[\gamma^2+\frac{v_*'^2}{v_*^2}(\mathcal{E}^2-\mathcal{E}M)\right],
	\end{align}
\end{subequations}
where $\Theta(x)$ is the Heaviside function.

In addition, we are also interested in the modified hybridization functions in the $J$ term calculations given by
\begin{subequations}\label{Eq:Delta_bar_Nf_4}
	\begin{align}
		\bar{\Delta}_1(\mathcal{E})=&\Theta(-\mathcal{E}-M)\frac{-\mathcal{E}}{8v_*^2}\Omega_0e^{-\frac{\lambda^2}{v_*^2}\left(\mathcal{E}^2+\mathcal{E}M\right)},\\
		\bar{\Delta}_2(\mathcal{E})=&\Theta(-\mathcal{E})\frac{-\mathcal{E}}{8v_*^2}\Omega_0e^{-\frac{\lambda^2}{v_*^2}\left(\mathcal{E}^2-\mathcal{E}M\right)},\\
		\bar{\Delta}_3(\mathcal{E})=&\Theta(\mathcal{E})\frac{\mathcal{E}}{8v_*^2}\Omega_0e^{-\frac{\lambda^2}{v_*^2}\left(\mathcal{E}^2+\mathcal{E}M\right)},\\
		\bar{\Delta}_4(\mathcal{E})=&\Theta(\mathcal{E}-M)\frac{\mathcal{E}}{8v_*^2}\Omega_0e^{-\frac{\lambda^2}{v_*^2}\left(\mathcal{E}^2-\mathcal{E}M\right)}.
	\end{align}
\end{subequations}

\subsection{$c$ fermion bands with a dynamical mass}\label{Sec:App:c_band_dyn_mass}
	
	In the local moment regime (i.e., $N_f$ at each site), an effective mass term due to the $\hat{H}_W$ term can be derived for $N_f\neq 4$. This interaction-induced mass term can modify the low-energy physics significantly. We focus on $N_f=1,2,3$ (corresponding to $\nu_f=-3,-2,-1$). $N_f=5,6,7$ can be obtained by performing a particle-hole transformation.
	
	The effective band structure with $\hat{W}$ term is determined by
	\begin{align}
		\nonumber&\hat{h}^{(\eta)}(\vex{q})+\hat{W}\\[2mm]
		=&\left[\begin{array}{cccc}
			0 & 0 & v_*|\vex{q}|e^{i\theta_{\eta,\vex{q}}} &0 \\
			0 & 0 &0 & v_*|\vex{q}|e^{-i\theta_{\eta,\vex{q}}} \\
			v_*|\vex{q}|e^{-i\theta_{\eta,\vex{q}}}& 0 &-\bar{W} &M \\
			0 & v_*|\vex{q}|e^{i\theta_{\eta,\vex{q}}} &M &-\bar{W} 
		\end{array}
		\right],
	\end{align}
	where $\bar{W}=-(N_f-4)(W_3-W_1)>M>0$. The expression of the energy bands are given by
	\begin{subequations}
		\begin{align}
			E_1(\vex{q})=&\frac{1}{2}\left(-M-\bar{W}-\sqrt{\left(M+\bar{W}\right)^2+4v_*^2|\vex{q}|^2}\right),\\
			E_2(\vex{q})=&\frac{1}{2}\left(M-\bar{W}-\sqrt{\left(M-\bar{W}\right)^2+4v_*^2|\vex{q}|^2}\right),\\
			E_3(\vex{q})=&\frac{1}{2}\left(-M-\bar{W}+\sqrt{\left(M+\bar{W}\right)^2+4v_*^2|\vex{q}|^2}\right),\\
			E_4(\vex{q})=&\frac{1}{2}\left(M-\bar{W}+\sqrt{\left(M-\bar{W}\right)^2+4v_*^2|\vex{q}|^2}\right).
		\end{align}
	\end{subequations}
	The above expressions are reduced to the single-particle results in Eq.~(\ref{Eq:App:E_c}) after setting $\bar{W}=0$.
	At $\Gamma$ point (i.e., $\vex{q}=0$), $E_1(0)=-M-\bar{W}$, $E_2(0)=M-\bar{W}$, $E_3(0)=0$, $E_4(0)=0$. The band structures with $\hat{W}$ show strong particle-hole asymmetry and a spectral gap $\bar{W}-M$. The corresponding wavefunctions take the same form as Eq.~(\ref{Eq:App:Psi_c}) with new $u_b$ and $v_b$ given by
	\begin{subequations}
		\begin{align}
			u_1(\vex{q})=&\frac{\frac{1}{2}\left(-M-\bar{W}\!+\!\sqrt{\left(M+\bar{W}\right)^2\!+\!4v_*^2|\vex{q}|^2}\right)e^{i\theta_{\eta,\vex{q}}}}{\sqrt{\frac{1}{4}\!\left(\!-M-\bar{W}\!+\!\sqrt{\left(M+\bar{W}\right)^2\!+\!4v_*^2|\vex{q}|^2}\right)^2\!\!+\!v_*^2|\vex{q}|^2}},\\
			v_1(\vex{q})=&\frac{v_*|\vex{q}|}{\sqrt{\frac{1}{4}\!\left(\!-M-\bar{W}\!+\!\sqrt{\left(M+\bar{W}\right)^2\!+\!4v_*^2|\vex{q}|^2}\right)^2\!\!+\!v_*^2|\vex{q}|^2}},\\
			u_2(\vex{q})=&\frac{\frac{1}{2}\left(M-\bar{W}\!+\!\sqrt{\left(M-\bar{W}\right)^2\!+\!4v_*^2|\vex{q}|^2}\right)e^{i\theta_{\eta,\vex{q}}}}{\sqrt{\frac{1}{4}\!\left(\!M-\bar{W}\!+\!\sqrt{\left(M-\bar{W}\right)^2\!+\!4v_*^2|\vex{q}|^2}\right)^2\!\!+\!v_*^2|\vex{q}|^2}},\\
			v_2(\vex{q})=&\frac{v_*|\vex{q}|}{\sqrt{\frac{1}{4}\!\left(\!M-\bar{W}\!+\!\sqrt{\left(M-\bar{W}\right)^2\!+\!4v_*^2|\vex{q}|^2}\right)^2\!\!+\!v_*^2|\vex{q}|^2}},\\
			u_3(\vex{q})=&\frac{\frac{1}{2}\left(M+\bar{W}\!+\!\sqrt{\left(M+\bar{W}\right)^2\!+\!4v_*^2|\vex{q}|^2}\right)e^{i\theta_{\eta,\vex{q}}}}{\sqrt{\frac{1}{4}\!\left(\!M+\bar{W}\!+\!\sqrt{\left(M+\bar{W}\right)^2\!+\!4v_*^2|\vex{q}|^2}\right)^2\!\!+\!v_*^2|\vex{q}|^2}},\\
			v_3(\vex{q})=&\frac{v_*|\vex{q}|}{\sqrt{\frac{1}{4}\!\left(\!M+\bar{W}\!+\!\sqrt{\left(M+\bar{W}\right)^2\!+\!4v_*^2|\vex{q}|^2}\right)^2\!\!+\!v_*^2|\vex{q}|^2}},\\
			u_4(\vex{q})=&\frac{\frac{1}{2}\left(-M+\bar{W}\!+\!\sqrt{\left(M-\bar{W}\right)^2\!+\!4v_*^2|\vex{q}|^2}\right)e^{i\theta_{\eta,\vex{q}}}}{\sqrt{\frac{1}{4}\!\left(\!-M+\bar{W}\!+\!\sqrt{\left(M-\bar{W}\right)^2\!+\!4v_*^2|\vex{q}|^2}\right)^2\!\!+\!v_*^2|\vex{q}|^2}},\\
			v_4(\vex{q})=&\frac{v_*|\vex{q}|}{\sqrt{\frac{1}{4}\!\left(\!-M+\bar{W}\!+\!\sqrt{\left(M-\bar{W}\right)^2\!+\!4v_*^2|\vex{q}|^2}\right)^2\!\!+\!v_*^2|\vex{q}|^2}}.
		\end{align}
	\end{subequations}
	Again, the above expressions are reduced to the single-particle results in Eq.~(\ref{Eq:App:Psi_c}) after setting $\bar{W}=0$.
	
	We also derive the density of states per spin per valley for each band as follows:
	\begin{subequations}
		\begin{align}
			\rho_1(\mathcal{E})=&\Theta(-\mathcal{E}-M-\bar{W})\frac{-2\mathcal{E}-M-\bar{W}}{4\pi v_*^2},\\
			\rho_2(\mathcal{E})=&\Theta(-\mathcal{E}+M-\bar{W})\frac{-2\mathcal{E}+M-\bar{W}}{4\pi v_*^2},\\
			\rho_3(\mathcal{E})=&\Theta(\mathcal{E})\frac{2\mathcal{E}+M+\bar{W}}{4\pi v_*^2},\\
			\rho_4(\mathcal{E})=&\Theta(\mathcal{E})\frac{2\mathcal{E}-M+\bar{W}}{4\pi v_*^2}.	
		\end{align}
	\end{subequations}
	It is worth mentioning that the total density of states at $E=0^+$ is significantly enhanced in the presence of the $\hat{W}$ term.

	The hybridization function between $f$ fermion and the $b$th band are given by 
	\begin{subequations}
		\begin{align}
			\nonumber\Delta_1(\mathcal{E})
			=&\Theta(-\mathcal{E}-M_+)\frac{-\mathcal{E}-M_+}{8v_*^2}\Omega_0e^{-\frac{\lambda^2}{v_*^2}(\mathcal{E}^2+\mathcal{E}M_+)}\\
			&\times\left[\gamma^2+\frac{v_*'^2}{v_*^2}(\mathcal{E}^2+\mathcal{E}M_+)\right],\\
			\nonumber\Delta_2(\mathcal{E})
			=&\Theta(-\mathcal{E}-|M_-|)\frac{-\mathcal{E}+M_-}{8v_*^2}\Omega_0e^{-\frac{\lambda^2}{v_*^2}(\mathcal{E}^2-\mathcal{E}M_-)}\\
			&\times\left[\gamma^2+\frac{v_*'^2}{v_*^2}(\mathcal{E}^2-\mathcal{E}M_-)\right],\\
			\nonumber\Delta_3(\mathcal{E})
			=&\Theta(\mathcal{E})\frac{\mathcal{E}+M_+}{8v_*^2}\Omega_0e^{-\frac{\lambda^2}{v_*^2}(\mathcal{E}^2+\mathcal{E}M_+)}\\
			&\times\left[\gamma^2+\frac{v_*'^2}{v_*^2}(\mathcal{E}^2+\mathcal{E}M_+)\right],\\		
			\nonumber\Delta_4(\mathcal{E})
			=&\Theta(\mathcal{E})\frac{\mathcal{E}-M_-}{8v_*^2}\Omega_0e^{-\frac{\lambda^2}{v_*^2}(\mathcal{E}^2-\mathcal{E}M_-)}\\
			&\times\left[\gamma^2+\frac{v_*'^2}{v_*^2}(\mathcal{E}^2-\mathcal{E}M_-)\right],
		\end{align}
	\end{subequations}
	where $M_{\pm}=M\pm\bar{W}$.
	
	The modified hybridization functions (associated with poor man's scaling of $J$) are given by
	\begin{subequations}
		\begin{align}
			\bar{\Delta}_1(\mathcal{E})=&\Theta(-\mathcal{E}-M_+)\frac{-\mathcal{E}}{8v_*^2}\Omega_0e^{-\frac{\lambda^2}{v_*^2}\left(\mathcal{E}^2+\mathcal{E}M_+\right)},\\
			\bar{\Delta}_2(\mathcal{E})=&\Theta(-\mathcal{E}-|M_-|)\frac{-\mathcal{E}}{8v_*^2}\Omega_0e^{-\frac{\lambda^2}{v_*^2}\left(\mathcal{E}^2-\mathcal{E}M_-\right)},\\
			\bar{\Delta}_3(\mathcal{E})=&\Theta(\mathcal{E})\frac{\mathcal{E}}{8v_*^2}\Omega_0e^{-\frac{\lambda^2}{v_*^2}\left(\mathcal{E}^2+\mathcal{E}M_+\right)},\\
			\bar{\Delta}_4(\mathcal{E})=&\Theta(\mathcal{E})\frac{\mathcal{E}}{8v_*^2}\Omega_0e^{-\frac{\lambda^2}{v_*^2}\left(\mathcal{E}^2-\mathcal{E}M_-\right)}.
		\end{align}
	\end{subequations}

	\begin{widetext}
	
	\section{Derivation of T-matrix corrections}\label{Sec:App:TMatrix}
	
	\subsection{Kondo coupling}
	
	The Kondo coupling can be expressed as follows:
	\begin{align}
		\hat{H}_{\text{imp},K}=&\frac{J_K}{\mathcal{N}_s\gamma^2}\!\!\sum_{\substack{\alpha,\eta,a,s\\ \alpha',\eta',a',s'}}\sum_{\vex{q},\vex{q}'}\left[V^{(\eta')}_{\alpha'a'}(\vex{q}')\right]^*V^{(\eta)}_{\alpha a}(\vex{q})\,\mathcal{S}_{\alpha,\eta,s;\alpha',\eta',s'} \left(c^{\dagger}_{\vex{q}',a',\eta',s'}c_{\vex{q},a,\eta,s}-\frac{1}{2}\delta_{(a,\eta,s),(a',\eta',s')}\delta_{\vex{q},\vex{q}'}\right)\!\hat{\mathcal{P}}_{N_f}.
	\end{align}
	where $\mathcal{S}_{\alpha,\eta,s;\alpha',\eta',s'}\equiv f^{\dagger}_{\alpha,\eta,s}f_{\alpha',\eta',s'}-\frac{1}{2}\delta_{(\alpha,\eta,s),(\alpha',\eta',s')}$ and $\delta_{(\alpha,\eta,s),(\alpha',\eta',s')}=\delta_{\alpha,\alpha'}\delta_{\eta,\eta'}\delta_{s,s'}$.
	
	\begin{align}
		\nonumber T^{(p)}(E)=&\frac{J_K^2}{\mathcal{N}_s^2}\sum_{\vex{q}_2,\vex{q}_1'}\sum_{\substack{\alpha_1,\eta_1,a_1,s_1\\ \alpha_1',\eta_1',a_1',s_1'}}\sum_{\substack{\alpha_2,\eta_2,a_2,s_2\\ \alpha_2',\eta_2',a_2',s_2'}}
		\frac{\left[V^{(\eta_1')}_{\alpha_1'a_1'}(\vex{q}_1')\right]^*V^{(\eta_2)}_{\alpha_2 a_2}(\vex{q}_2)}{\gamma^2}
		c^{\dagger}_{\vex{q}_1',a_1',\eta_1',s_1'}c_{\vex{q}_2,a_2,\eta_2,s_2}\mathcal{S}_{\alpha_1,\eta_1,s_1;\alpha_1',\eta_1',s_1'}\mathcal{S}_{\alpha_2,\eta_2,s_2;\alpha_2',\eta_2',s_2'}\hat{\mathcal{P}}_{N_f}\\
		&\times\sum_{b}\sum_{\vex{p}}'\frac{\left[V^{(\eta_1)}_{\alpha_1a_1}(\vex{p})\right]^*V^{(\eta_2')}_{\alpha_2' a_2'}(\vex{p})}{\gamma^2}\frac{\Psi^{(\eta_1)}_{b,a_1}(\vex{p})\left[\Psi^{(\eta_2')}_{b,a_2'}(\vex{p})\right]^*}{E-E_b(\vex{p})+\mu_c}\delta_{\eta_1,\eta_2'}\delta_{s_1,s_2'},\\
		\nonumber T^{(h)}(E)=&\frac{J_K^2}{\mathcal{N}_s^2}\sum_{\vex{q}_1,\vex{q}_2'}\sum_{\substack{\alpha_1,\eta_1,a_1,s_1\\ \alpha_1',\eta_1',a_1',s_1'}}\sum_{\substack{\alpha_2,\eta_2,a_2,s_2\\ \alpha_2',\eta_2',a_2',s_2'}}
		\frac{V^{(\eta_1)}_{\alpha_1a_1}(\vex{q}_1)\left[V^{(\eta_2')}_{\alpha_2' a_2'}(\vex{q}_2')\right]^*}{\gamma^2}
		c_{\vex{q}_1,a_1,\eta_1,s_1}c^{\dagger}_{\vex{q}_2',a_2',\eta_2',s_2'}\mathcal{S}_{\alpha_1,\eta_1,s_1;\alpha_1',\eta_1',s_1'}\mathcal{S}_{\alpha_2,\eta_2,s_2;\alpha_2',\eta_2',s_2'}\hat{\mathcal{P}}_{N_f}\\
		&\times\sum_{b}\sum_{\vex{p}}''\frac{\left[V^{(\eta_1')}_{\alpha_1' a_1'}(\vex{p})\right]^*V^{(\eta_2)}_{\alpha_2a_2}(\vex{p})}{\gamma^2}\frac{\left[\Psi^{(\eta_1')}_{b,a_1'}(\vex{p})\right]^*\Psi^{(\eta_2)}_{b,a_2}(\vex{p})}{E+E_b(\vex{p})-\mu_c}\delta_{\eta_1',\eta_2}\delta_{s_1',s_2},
	\end{align}
	where $\sum_{\vex{p}}'$ sums over all the momenta with $D-\delta D<E_b(\vex{p})-\mu_c<D$ and $\sum_{\vex{p}}''$ sums over all the momenta with $-D<E_b(\vex{p})-\mu_c<-D+\delta D$. 
	In the above expressions, the Kronecker delta functions on valley and spin sectors are due to the valley and spin conservation in the $c$ fermions. The orbital degrees of freedom are not unconstrained. As a result, we need to separate the \textit{internal} summations $\sum_{\alpha,\alpha'}$ into $\sum_{\alpha,\alpha'}\left(\delta_{\alpha,\alpha'}+\delta_{\bar{\alpha},\alpha'}\right)$, where $\bar{1}=2$ and $\bar{2}=1$. The equal-orbital contributions give rise to corrections to the Kondo coupling and impurity potential scattering terms. On the other hand, the opposite-orbital contributions generate additional terms that are not important for our analysis, and these contributions can be ignored after summing over the band indices as long as $D\gg \mu_c, m$. Thus, we focus on the equal-orbital contributions in the T-matrix calculations. Thus, the internal summations involve the following identities:
	\begin{align}
		&\sum_{\nu'}f^{\dagger}_{\nu'}f_{\nu_1}f^{\dagger}_{\nu_2}f_{\nu'}\hat{\mathcal{P}}_{N_f}=\left[N_f\delta_{\nu_1,\nu_2}+(1-N_f)f^{\dagger}_{\nu_2}f_{\nu_1}\right]\hat{\mathcal{P}}_{N_f},\\
		\rightarrow&\sum_{\nu'}\mathcal{S}_{\nu';\nu_1}\mathcal{S}_{\nu_2;\nu'}\hat{\mathcal{P}}_{N_f}=\left[\left(\frac{N_f}{2}+\frac{1}{4}\right)\delta_{\nu_1,\nu_2}-N_f\mathcal{S}_{\nu_2;\nu_1}\right]\hat{\mathcal{P}}_{N_f},\\
		&\sum_{\nu'}f^{\dagger}_{\nu_1}f_{\nu'}f^{\dagger}_{\nu'}f_{\nu_2}\hat{\mathcal{P}}_{N_f}=\left[(N-N_f+1)f^{\dagger}_{\nu_1}f_{\nu_2}\right]\hat{\mathcal{P}}_{N_f},\\
		\rightarrow&\sum_{\nu'}\mathcal{S}_{\nu_1;\nu'}\mathcal{S}_{\nu';\nu_2}\hat{\mathcal{P}}_{N_f}=\left[\left(\frac{N-N_f}{2}+\frac{1}{4}\right)\delta_{\nu_1,\nu_2}+(N-N_f)\mathcal{S}_{\nu_1;\nu_2}\right]\hat{\mathcal{P}}_{N_f},
	\end{align}
	where $\nu$ is the shorthand notation for $(\alpha,\eta,s)$ and $N=8$. Lastly, the angular average eliminates the mixing between $\gamma$ and $v_*'$ terms, and the calculations are further simplified.

	With the ideas mentioned above, the T-matrix contributions (with $E=0$) become
	\begin{align}
		\nonumber &T^{(p)}(E=0)\\
		\nonumber\approx&\frac{J_K^2}{\mathcal{N}_s}\left(\frac{\Omega_0\delta D}{-D}\right)\sum_{\vex{q}_2,\vex{q}_1'}\sum_{\substack{\alpha_2,\eta_2,a_2,s_2\\ \alpha_1',\eta_1',a_1',s_1'}}
		\frac{\left[V^{(\eta_1')}_{\alpha_1'a_1'}(\vex{q}_1')\right]^*V^{(\eta_2)}_{\alpha_2 a_2}(\vex{q}_2)}{\gamma^2}
		c^{\dagger}_{\vex{q}_1',a_1',\eta_1',s_1'}c_{\vex{q}_2,a_2,\eta_2,s_2}\\
		&\times\left[-N_f\mathcal{S}_{\alpha_2,\eta_2,s_2;\alpha_1',\eta_1',s_1'}+\left(\frac{N_f}{2}+\frac{1}{4}\right)\delta_{(\alpha_1',\eta_1',s_1'),(\alpha_2,\eta_2,s_2)}\right]
		\sum_{b}\rho_b(D+\mu_c)\left[\frac{\gamma^2+v_*'^2|\vex{p}|^2}{\gamma^2}\left|u_b(\vex{p})\right|^2\right]\bigg|_{E_b(\vex{p})-\mu_c=D},\\
		=&\left[\frac{N_f J_K^2}{\pi}\frac{\delta D}{D}\sum_b\frac{\Delta_b(D+\mu_c)}{\gamma^2}\right]\frac{1}{\mathcal{N}_s}\sum_{\vex{q}_2,\vex{q}_1'}\sum_{\substack{\alpha_2,\eta_2,a_2,s_2\\ \alpha_1',\eta_1',a_1',s_1'}}
		\frac{\left[V^{(\eta_1')}_{\alpha_1'a_1'}(\vex{q}_1')\right]^*V^{(\eta_2)}_{\alpha_2 a_2}(\vex{q}_2)}{\gamma^2}
		c^{\dagger}_{\vex{q}_1',a_1',\eta_1',s_1'}c_{\vex{q}_2,a_2,\eta_2,s_2}\mathcal{S}_{\alpha_2,\eta_2,s_2;\alpha_1',\eta_1',s_1'}+\dots,\\
		\nonumber &T^{(h)}(E=0)\\
		\nonumber\approx&-\frac{J_K^2}{\mathcal{N}_s}\left(\frac{\Omega_0\delta D}{-D}\right)\sum_{\vex{q}_1,\vex{q}_2'}\sum_{\substack{\alpha_1,\eta_1,a_1,s_1\\ \alpha_2',\eta_2',a_2',s_2'}}
		\frac{V^{(\eta_1)}_{\alpha_1a_1}(\vex{q}_1)\left[V^{(\eta_2')}_{\alpha_2' a_2'}(\vex{q}_2')\right]^*}{\gamma^2}
		c^{\dagger}_{\vex{q}_2',a_2',\eta_2',s_2'}c_{\vex{q}_1,a_1,\eta_1,s_1}\\
		&\times\left[\left(N-N_f\right)\mathcal{S}_{\alpha_1,\eta_1,s_1;\alpha_2',\eta_2',s_2'}+\left(\frac{N-N_f}{2}+\frac{1}{4}\right)\delta_{(\alpha_1,\eta_1,s_1),(\alpha_2',\eta_2',s_2')}\right]\sum_{b}\rho_b(-D+\mu_c)\left[\frac{\gamma^2+v_*'^2|\vex{p}|^2}{\gamma^2}\left|u_b(\vex{p})\right|^2\right]\bigg|_{E_b(\vex{p})-\mu_c=-D}\\
		=&\left[\frac{\left(N-N_f\right) J_K^2}{\pi}\frac{\delta D}{D}\sum_b\frac{\Delta_b(-D+\mu_c)}{\gamma^2}\right]\frac{1}{\mathcal{N}_s}\sum_{\vex{q}_1,\vex{q}_2'}\sum_{\substack{\alpha_1,\eta_1,a_1,s_1\\ \alpha_2',\eta_2',a_2',s_2'}}
		\frac{V^{(\eta_1)}_{\alpha_1a_1}(\vex{q}_1)\left[V^{(\eta_2')}_{\alpha_2' a_2'}(\vex{q}_2')\right]^*}{\gamma^2}
		c^{\dagger}_{\vex{q}_2',a_2',\eta_2',s_2'}c_{\vex{q}_1,a_1,\eta_1,s_1}\mathcal{S}_{\alpha_1,\eta_1,s_1;\alpha_2',\eta_2',s_2'}+\dots,
	\end{align}
	where we have ignored the potential scattering terms in the final expressions. The minus sign in $T^{(h)}$ is due to swapping the fermionic operators.
	
	\subsection{Hund's rule coupling}
	
	We express the impurity Hund's rule coupling as follows:
	\begin{align}
		\hat{H}_{\text{imp},J}=&-\frac{1}{\mathcal{N}_s}\sum_{\eta,\alpha,s_1,s_2}\sum_{\vex{q},\vex{q}'}e^{-\frac{\lambda^2\left(|\vex{q}|^2+|\vex{q}'|^2\right)}{2}}\left[
		\begin{array}{c}
			J_1\mathcal{S}_{\alpha,\eta,s_1;\alpha,\eta,s_2}:c^{\dagger}_{\vex{q}',\alpha+2,\eta,s_2}c_{\vex{q},\alpha+2,\eta,s_1}:\\[2mm]
			-J_2\mathcal{S}_{\bar{\alpha},-\eta,s_1;\alpha,\eta,s_2}c^{\dagger}_{\vex{q}',\alpha+2,\eta,s_2}c_{\vex{q},\bar{\alpha}+2,-\eta,s_1}
		\end{array}
		\right],
	\end{align}
where we will set $J_1=J_2=J$ at the end of calculations. The $J_1$ and $J_2$ terms describe two distinct processes in the perturbation theory. From the symmetry point of view, it is sufficient to focus on the renormalization of $J_1$ term as the entire Hund's rule coupling can be expressed as a $U(4)$ interaction. Thus, we focus only on $J_1$ term in this section.
	
	At the one-loop level, the T-matrix contributions to the $J_1$ term are as follows:
	\begin{align}
		\nonumber \bar{T}^{(p)}(E)=&\frac{J_1^2}{\mathcal{N}_s^2}\sum_{\vex{q},\vex{q}'}\sum_{\substack{\alpha_1,\eta_1,s_1,s_1'\\ \alpha_2,\eta_2,s_2,s_2'}}e^{-\frac{\lambda^2}{2}(|\vex{q}|^2+|\vex{q}'|^2)}c^{\dagger}_{\vex{q}',\alpha_1+2,\eta_1,s_1'}c_{\vex{q},\alpha_2+2,\eta_2,s_2}\mathcal{S}_{\alpha_1,\eta_1,s_1;\alpha_1,\eta_1,s_1'}\mathcal{S}_{\alpha_2,\eta_2,s_2;\alpha_2,\eta_2,s_2'}\\
		\nonumber &\times\sum_{b}\sum_{\vex{p}}'e^{-\lambda^2|\vex{p}|^2}\frac{\psi^{(\eta_1)}_{b,\alpha_1+2}(\vex{p})\left[\psi^{(\eta_2)}_{b,\alpha_2+2}(\vex{p})\right]^*}{E-E_b(\vex{p})+\mu_c}\delta_{\eta_1,\eta_2}\delta_{s_1,s_2'}\\
		\nonumber &+\frac{J_2^2}{\mathcal{N}_s^2}\sum_{\vex{q},\vex{q}'}\sum_{\substack{\alpha_1,\eta_1,s_1,s_1'\\ \alpha_2,\eta_2,s_2,s_2'}}e^{-\frac{\lambda^2}{2}(|\vex{q}|^2+|\vex{q}'|^2)}c^{\dagger}_{\vex{q}',\alpha_1+2,\eta_1,s_1'}c_{\vex{q},\bar\alpha_2+2,-\eta_2,s_2}\mathcal{S}_{\bar\alpha_1,-\eta_1,s_1;\alpha_1,\eta_1,s_1'}\mathcal{S}_{\bar\alpha_2,-\eta_2,s_2;\alpha_2,\eta_2,s_2'}\\
		&\times\sum_{b}\sum_{\vex{p}}'e^{-\lambda^2|\vex{p}|^2}\frac{\psi^{(b,-\eta_1)}_{\bar\alpha_1+2}(\vex{p})\left[\psi^{(b,\eta_2)}_{\alpha_2+2}(\vex{p})\right]^*}{E-E_b(\vex{p})+\mu_c}\delta_{-\eta_1,\eta_2}\delta_{s_1,s_2'},\\
		\nonumber \bar{T}^{(h)}(E)=&\frac{J_1^2}{\mathcal{N}_s^2}\sum_{\vex{q},\vex{q}'}\sum_{\substack{\alpha_1,\eta_1,s_1,s_1'\\ \alpha_2,\eta_2,s_2,s_2'}}e^{-\frac{\lambda^2}{2}(|\vex{q}|^2+|\vex{q}'|^2)}c_{\vex{q}',\alpha_1+2,\eta_1,s_1}c^{\dagger}_{\vex{q},\alpha_2+2,\eta_2,s_2'}\mathcal{S}_{\alpha_1,\eta_1,s_1;\alpha_1,\eta_1,s_1'}\mathcal{S}_{\alpha_2,\eta_2,s_2;\alpha_2,\eta_2,s_2'}\\
		\nonumber &\times\sum_{b}\sum_{\vex{p}}''e^{-\lambda^2|\vex{p}|^2}\frac{\left[\psi^{(\eta_1)}_{b,\alpha_1+2}(\vex{p})\right]^*\psi^{(\eta_2)}_{b,\alpha_2+2}(\vex{p})}{E+E_b(\vex{p})-\mu_c}\delta_{\eta_1,\eta_2}\delta_{s_1',s_2}\\
		\nonumber &+\frac{J_2^2}{\mathcal{N}_s^2}\sum_{\vex{q},\vex{q}'}\sum_{\substack{\alpha_1,\eta_1,s_1,s_1'\\ \alpha_2,\eta_2,s_2,s_2'}}e^{-\frac{\lambda^2}{2}(|\vex{q}|^2+|\vex{q}'|^2)}c_{\vex{q}',\bar\alpha_1+2,-\eta_1,s_1}c^{\dagger}_{\vex{q},\alpha_2+2,\eta_2,s_2'}\mathcal{S}_{\bar\alpha_1,-\eta_1,s_1;\alpha_1,\eta_1,s_1'}\mathcal{S}_{\bar\alpha_2,-\eta_2,s_2;\alpha_2,\eta_2,s_2'}\\
		&\times\sum_{b}\sum_{\vex{p}}''e^{-\lambda^2|\vex{p}|^2}\frac{\left[\psi^{(\eta_1)}_{b,\bar\alpha_1+2}(\vex{p})\right]^*\psi^{(-\eta_2)}_{b,\alpha_2+2}(\vex{p})}{E+E_b(\vex{p})-\mu_c}\delta_{\eta_1,-\eta_2}\delta_{s_1',s_2},
	\end{align}
	where $\sum_{\vex{p}}'$ sums over all the momenta with $D-\delta D<E_b(\vex{p})-\mu_c<D$ and $\sum_{\vex{p}}''$ sums over all the momenta with $-D<E_b(\vex{p})-\mu_c<-D+\delta D$. 
	Similar to the Kondo case, we retain only the equal-orbital contributions.

	With the ideas mentioned above, the T-matrix contributions become
	\begin{align}
		\nonumber \bar{T}^{(p)}(E=0)\approx&\frac{(J_1^2+J_2^2)}{\mathcal{N}_s}\left(\frac{\Omega_0\delta D}{-D}\right)\sum_{\vex{q},\vex{q}'}\sum_{\alpha_1,\eta_1,s_2,s_1'}e^{-\frac{\lambda^2}{2}(|\vex{q}|^2+|\vex{q}'|^2)}c^{\dagger}_{\vex{q}',\alpha_1+2,\eta_1,s_1'}c_{\vex{q},\alpha_1+2,\eta_1,s_2}\\
		&\times\sum_{s_1}\mathcal{S}_{\alpha_1,\eta_1,s_1;\alpha_1,\eta_1,s_1'}\mathcal{S}_{\alpha_1,\eta_1,s_2;\alpha_1,\eta_1,s_1}\sum_{b}\rho_b(D+\mu_c)\left[e^{-\lambda^2|\vex{p}|^2}\left|\psi^{(\eta_1)}_{b,\alpha_1+2}(\vex{p})\right|^2\right]\bigg|_{E_b(\vex{p})-\mu_c=D},\\
		\nonumber \bar{T}^{(h)}(E=0)\approx&-\frac{(J_1^2+J_2^2)}{\mathcal{N}_s}\left(\frac{\Omega_0\delta D}{-D}\right)\sum_{\vex{q},\vex{q}'}\sum_{\alpha_1,\eta_1,s_1,s_2'}e^{-\frac{\lambda^2}{2}(|\vex{q}|^2+|\vex{q}'|^2)}c^{\dagger}_{\vex{q},\alpha_1+2,\eta_1,s_2'}c_{\vex{q}',\alpha_1+2,\eta_1,s_1}\\
		&\times\sum_{s_1'}\mathcal{S}_{\alpha_1,\eta_1,s_1;\alpha_1,\eta_1,s_1'}\mathcal{S}_{\alpha_1,\eta_2,s_1';\alpha_2,\eta_2,s_2'}
		\sum_{b}\rho_b(-D+\mu_c)\left[e^{-\lambda^2|\vex{p}|^2}\left|\psi^{(\eta_1)}_{b,\alpha_1+2}(\vex{p})\right|^2\right]\bigg|_{E_b(\vex{p})-\mu_c=-D}.
	\end{align}

The internal summations are given by
\begin{align}
	\sum_{s'}f^{\dagger}_{\alpha,\eta,s'}f_{\alpha,\eta,s_1'}f^{\dagger}_{\alpha,\eta,s_2}f_{\alpha,\eta,s'}=&\left(1-\sum_{s'}f^{\dagger}_{\alpha_1,\eta_1,s_1}f_{\alpha_1,\eta_1,s_1}\right)f^{\dagger}_{\alpha_1,\eta_1,s_2}f_{\alpha_1,\eta_1,s_1'}+\delta_{s_1',s_2}\left(\sum_{s'}f^{\dagger}_{\alpha_1,\eta_1,s_1}f_{\alpha_1,\eta_1,s_1}\right),\\
	\sum_{s'}\mathcal{S}_{\alpha,\eta,s';\alpha,\eta,s_1'}\mathcal{S}_{\alpha,\eta,s_2;\alpha,\eta,s'}=&-\left(\sum_{s'}f^{\dagger}_{\alpha_1,\eta_1,s_1}f_{\alpha_1,\eta_1,s_1}\right)\mathcal{S}_{\alpha,\eta,s_2;\alpha,\eta,s_1'}+\delta_{s_1',s_2}\left(\frac{1}{4}+\frac{1}{2}\sum_{s'}f^{\dagger}_{\alpha_1,\eta_1,s_1}f_{\alpha_1,\eta_1,s_1}\right),\\
	\sum_{s'}f^{\dagger}_{\alpha,\eta,s_2}f_{\alpha,\eta,s'}f^{\dagger}_{\alpha,\eta,s'}f_{\alpha,\eta,s_1'}=&f^{\dagger}_{\alpha_1,\eta_1,s_2}f_{\alpha_1,\eta_1,s_1'}\left(3-\sum_{s'}f^{\dagger}_{\alpha_1,\eta_1,s_1}f_{\alpha_1,\eta_1,s_1}\right),\\
	\sum_{s'}\mathcal{S}_{\alpha,\eta,s';\alpha,\eta,s_1'}\mathcal{S}_{\alpha,\eta,s_2;\alpha,\eta,s'}=&\mathcal{S}_{\alpha,\eta,s_2;\alpha,\eta,s_1'}\left(2-\sum_{s'}f^{\dagger}_{\alpha_1,\eta_1,s_1}f_{\alpha_1,\eta_1,s_1}\right)+\delta_{s_1',s_2}\left(\frac{5}{4}-\frac{1}{2}\sum_{s'}f^{\dagger}_{\alpha_1,\eta_1,s_1}f_{\alpha_1,\eta_1,s_1}\right),
\end{align}
where we ignored terms associated with potential scatterings in the equations with $\mathcal{S}$.
Unlike the Kondo coupling calculations, the summations of products of two $\mathcal{S}$ are rather complicated because of  $\sum_{s'}f^{\dagger}_{\alpha_1,\eta_1,s_1}f_{\alpha_1,\eta_1,s_1}$.

\subsubsection{Particle-hole symmetric case}

	In a special limit that $\mu_c=0$ and $\rho(D)=\rho(-D)$, the expression can be simplified using the Hubbard operator identity \cite{Coleman2015introduction},
	\begin{align}
		\nonumber&\sum_{s'}\left[\mathcal{S}_{\alpha,\eta,s';\alpha,\eta,s_1'},\mathcal{S}_{\alpha,\eta,s_2;\alpha,\eta,s'}\right]=\sum_{s'}\left[f^{\dagger}_{\alpha,\eta,s'}f_{\alpha,\eta,s_1'},f^{\dagger}_{\alpha,\eta,s_2}f_{\alpha,\eta,s'}\right]\\
		=&\sum_{s'}f^{\dagger}_{\alpha,\eta,s'}f_{\alpha,\eta,s'}\delta_{s_1',s_2}-2f^{\dagger}_{\alpha,\eta,s_2}f_{\alpha,\eta,s_1'}=-2\mathcal{S}_{\alpha,\eta,s_2;\alpha,\eta,s_1'}.
	\end{align}
	With the Hubbard operator identity, we obtain 
	\begin{align}
		\nonumber\bar{T}^{(p)}+\bar{T}^{(h)}\approx&2\frac{(J_1^2+J_2^2)}{\mathcal{N}_s}\left(\frac{\Omega_0\delta D}{D}\right)\sum_{\vex{q},\vex{q}'}\sum_{\alpha_1,\eta_1,s_2,s_1'}e^{-\frac{\lambda^2}{2}(|\vex{q}|^2+|\vex{q}'|^2)}c^{\dagger}_{\vex{q}',\alpha_1+2,\eta_1,s_1'}c_{\vex{q},\alpha_1+2,\eta_1,s_2}\mathcal{S}_{\alpha_1,\eta_1,s_2;\alpha_1+2,\eta_1,s_1'}\\
		&\times\sum_{b}\rho_b(D)\left[e^{-\lambda^2|\vex{p}|^2}\left|\psi^{(\eta_1)}_{b,\alpha_1+2}(\vex{p})\right|^2\right]\bigg|_{E_b(\vex{p})=D}.
	\end{align}
	Setting $J_1=J_2=J$, we obtain
	\begin{align}
		J(D-\delta D)=J(D)-4J^2\delta D\sum_{b=3,4}\frac{\bar{\Delta}_b(D)}{\pi D},
	\end{align}	
	where $\bar{\Delta}_b$ is the modified hybridization function.
	
	Thus, the poor man's scaling equation for $J$ is given by
	\begin{align}
		\frac{dJ}{dD}=4J^2\sum_{b=3,4}\frac{\bar{\Delta}_b(D)}{\pi D}=\frac{J^2}{2\pi v_*^2}
		\left[e^{-\frac{\lambda^2}{v_*^2}\left(D^2+DM\right)}+\Theta(D-M)e^{-\frac{\lambda^2}{v_*^2}\left(D^2-DM\right)}\right]
	\end{align}
	
	\subsection{General case}
	
		For the general case, we surmise that $\sum_{s'}f^{\dagger}_{\alpha_1,\eta_1,s_1}f_{\alpha_1,\eta_1,s_1}$ can be replaced by the expectation value, i.e., $\left\langle\sum_{s'}f^{\dagger}_{\alpha_1,\eta_1,s_1}f_{\alpha_1,\eta_1,s_1}\right\rangle=N_f/4$. Thus,
		\begin{align}
			\sum_{s'}\mathcal{S}_{\alpha,\eta,s';\alpha,\eta,s_1'}\mathcal{S}_{\alpha,\eta,s_2;\alpha,\eta,s'}\approx&-\frac{N_f}{4}\mathcal{S}_{\alpha,\eta,s_2;\alpha,\eta,s_1'}+\left(\frac{N_f}{8}+\frac{1}{4}\right)\delta_{s_2,s_1'},\\
			\sum_{s'}\mathcal{S}_{\alpha,\eta,s';\alpha,\eta,s_1'}\mathcal{S}_{\alpha,\eta,s_2;\alpha,\eta,s'}\approx&\frac{8-N_f}{4}\mathcal{S}_{\alpha,\eta,s_2;\alpha,\eta,s_1'}+\left(\frac{8-N_f}{8}+\frac{1}{4}\right)\delta_{s_2,s_1'},
		\end{align}

		The T-matrix contributions are approximated by
	\begin{align}
		&\nonumber \bar{T}^{(p)}(E=0)\\
		\nonumber\approx&\frac{(J_1^2+J_2^2)}{\mathcal{N}_s}\left(\frac{\Omega_0\delta D}{-D}\right)\sum_{\vex{q},\vex{q}'}\sum_{\alpha_1,\eta_1,s_2,s_1'}e^{-\frac{\lambda^2}{2}(|\vex{q}|^2+|\vex{q}'|^2)}c^{\dagger}_{\vex{q}',\alpha_1+2,\eta_1,s_1'}c_{\vex{q},\alpha_1+2,\eta_1,s_2}\\
		\nonumber&\times\left(-\frac{N_f}{4}\mathcal{S}_{\alpha_1,\eta_1,s_2;\alpha_1,\eta_1,s_1'}+\frac{2+N_f}{8}\delta_{s_2,s_1'}\right)\sum_{b}\rho_b(D+\mu_c)\left[e^{-\lambda^2|\vex{p}|^2}\left|\psi^{(\eta_1)}_{b,\alpha_1+2}(\vex{p})\right|^2\right]\bigg|_{E_b(\vex{p})-\mu_c=D},\\
		=&\left[\frac{N_f}{4}\frac{J_1^2+J_2^2}{\pi}\frac{\delta D}{ D}\sum_b\bar{\Delta}_b(D+\mu_c)\right]\frac{1}{\mathcal{N}_s}\sum_{\vex{q},\vex{q}'}\sum_{\alpha_1,\eta_1,s_2,s_1'}\!\!e^{-\frac{\lambda^2}{2}(|\vex{q}|^2+|\vex{q}'|^2)}c^{\dagger}_{\vex{q}',\alpha_1+2,\eta_1,s_1'}c_{\vex{q},\alpha_1+2,\eta_1,s_2}\mathcal{S}_{\alpha_1,\eta_1,s_2;\alpha_1,\eta_1,s_1'}+\dots,\\
		\nonumber &\bar{T}^{(h)}(E=0)\\
		\nonumber\approx&-\frac{(J_1^2+J_2^2)}{\mathcal{N}_s}\left(\frac{\Omega_0\delta D}{-D}\right)\sum_{\vex{q},\vex{q}'}\sum_{\alpha_1,\eta_1,s_1,s_2'}e^{-\frac{\lambda^2}{2}(|\vex{q}|^2+|\vex{q}'|^2)}c^{\dagger}_{\vex{q},\alpha_1+2,\eta_1,s_2'}c_{\vex{q}',\alpha_1+2,\eta_1,s_1}\\
		\nonumber&\times\left(\frac{8-N_f}{4}\mathcal{S}_{\alpha_1,\eta_1,s_1;\alpha_1,\eta_1,s_2'}+\frac{10-N_f}{8}\delta_{s_1,s_2'}\right)
		\sum_{b}\rho_b(-D+\mu_c)\left[e^{-\lambda^2|\vex{p}|^2}\left|\psi^{(\eta_1)}_{b,\alpha_1+2}(\vex{p})\right|^2\right]\bigg|_{E_b(\vex{p})-\mu_c=-D}\\
		=&\left[\frac{8-N_f}{4}\frac{J_1^2+J_2^2}{\pi}\frac{\delta D}{ D}\sum_b\bar{\Delta}_b(-D+\mu_c)\right]\frac{1}{\mathcal{N}_s}\sum_{\vex{q},\vex{q}'}\sum_{\alpha_1,\eta_1,s_1,s_2'}e^{-\frac{\lambda^2}{2}(|\vex{q}|^2+|\vex{q}'|^2)}c^{\dagger}_{\vex{q},\alpha_1+2,\eta_1,s_2'}c_{\vex{q}',\alpha_1+2,\eta_1,s_1}\mathcal{S}_{\alpha_1,\eta_1,s_1;\alpha_1,\eta_1,s_2'}+\dots,
	\end{align}
	where we have ignored the potential scattering term in the final expression.

	Setting $J_1=J_2=J$, we obtain
	\begin{align}
		J(D-\delta D)=J(D)-\frac{N_f}{2}\frac{J^2(D)}{\pi}\frac{\delta D}{ D}\sum_b\bar{\Delta}_b(D+\mu_c)-\left(4-\frac{N_f}{2}\right)\frac{J^2(D)}{\pi}\frac{\delta D}{ D}\sum_b\bar{\Delta}_b(-D+\mu_c).
	\end{align}
	When $\mu_c=0$ and $N_f=4$, this reproduces the exact particle-hole symmetric result. The poor man's scaling equation is given by
	\begin{align}
		\frac{dJ}{dD}=\frac{N_f}{2}J^2\sum_b\frac{\bar{\Delta}_b(D+\mu_c)}{\pi D}+\left(4-\frac{N_f}{2}\right)J^2\sum_b\frac{\bar{\Delta}_b(-D+\mu_c)}{\pi D}
	\end{align}

\section{Impurity scattering terms}\label{Sec:App:Imp}
	
	Based on the T-matrix calculations, impurity potential scattering terms are generated. The potential scattering terms from $J_K$ and $J$ are distinct and correspond to orbital sector $a=1,2$ and $a=3,4$, respectively. The impact of the impurity potential scatterings on MATBG is beyond the purpose of this work. For completeness, we summarize the results in this section.
	
	The potential scattering terms from $T^{(p)}(E=0)+T^{(h)}(E=0)$ is given by
	\begin{align}
		\nonumber\delta \hat{H}_{\text{imp},\delta}=&\left\{-\left(\frac{N_f}{2}+\frac{1}{4}\right) J_K^2\delta D\left[\sum_b\frac{\Delta_b(D+\mu_c)}{\pi D\gamma^2}\right]+\left(\frac{N-N_f}{2}+\frac{1}{4}\right)J_K^2\delta D\left[\sum_b\frac{\Delta_b(-D+\mu_c)}{\pi D\gamma^2}\right]\right\}\\
		&\times\frac{1}{\mathcal{N}_s}\sum_{\vex{q},\vex{q}'}\sum_{\alpha,\eta,s}\sum_{a,a'}
		\frac{\left[V^{(\eta)}_{\alpha a'}(\vex{q}')\right]^*V^{(\eta)}_{\alpha a}(\vex{q})}{\gamma^2}
		c^{\dagger}_{\vex{q}',a',\eta,s}c_{\vex{q},a,\eta,s}\hat{\mathcal{P}}_{N_f}.
	\end{align} 
	Note that $\delta \hat{H}_{\text{imp},\delta}$ vanishes exactly when $N_f=4$ and $\mu_c=0$.
	
	The potential scattering terms from $\bar{T}^{(p)}(E=0)+\bar{T}^{(h)}(E=0)$ is given by	
	\begin{align}
			\nonumber\delta \hat{H}_{\text{imp},\delta'}=&\left\{-\left(\frac{N_f}{8}+\frac{1}{4}\right)(J_1^2+J_2^2)\delta D\left[\sum_b\frac{\bar{\Delta}_b(D+\mu_c)}{\pi D}\right]
			+\left(\frac{8-N_f}{8}+\frac{1}{4}\right)(J_1^2+J_2^2)\delta D\left[\sum_b\frac{\bar{\Delta}_b(-D+\mu_c)}{\pi D}\right]
			\right\}\\
			&\times\frac{1}{\mathcal{N}_s}\sum_{\vex{q},\vex{q}'}\sum_{\alpha,\eta,s,s}\!\!e^{-\frac{\lambda^2}{2}(|\vex{q}|^2+|\vex{q}'|^2)}c^{\dagger}_{\vex{q}',\alpha+2,\eta,s}c_{\vex{q},\alpha+2,\eta,s}
	\end{align}
	Again, $\delta \hat{H}_{\text{imp},\delta'}$ vanishes exactly when $N_f=4$ and $\mu_c=0$. We focus only on $J_1=J_2=J$. The potential scattering due to $J$ can be described by
		\begin{align}\label{Eq:H_P'}
	 \hat{H}_{\text{imp},\delta'}=\frac{\delta'}{\mathcal{N}_s}\sum_{\vex{q},\vex{q}'}\sum_{\alpha,\eta,s,s}\!\!e^{-\frac{\lambda^2}{2}(|\vex{q}|^2+|\vex{q}'|^2)}c^{\dagger}_{\vex{q}',\alpha+2,\eta,s}c_{\vex{q},\alpha+2,\eta,s},
	\end{align}	
	where the bare value of $\delta'$ is set to zero.
	
	The one-loop scaling flows are given by
	\begin{subequations}\label{Eq:ddelta_dD}
			\begin{align}
			\frac{d\delta}{dD}=&\left(\frac{N_f}{2}+\frac{1}{4}\right) J_K^2\left[\sum_b\frac{\Delta_b(D+\mu_c)}{\pi D\gamma^2}\right]-\left(\frac{N-N_f}{2}+\frac{1}{4}\right)J_K^2\left[\sum_b\frac{\Delta_b(-D+\mu_c)}{\pi D\gamma^2}\right],\\
			\frac{d\delta'}{dD}=&\left(\frac{N_f}{4}+\frac{1}{2}\right)J^2\left[\sum_b\frac{\bar{\Delta}_b(D+\mu_c)}{\pi D}\right]-\left(\frac{8-N_f}{4}+\frac{1}{2}\right)J^2\left[\sum_b\frac{\bar{\Delta}_b(-D+\mu_c)}{\pi D}\right].
		\end{align}
	\end{subequations}

	Since $J_K^2$ is typically much larger than $J^2$, $|\delta|$ is parametrically larger than $|\delta'|$ in general. This cause an imbalance in the orbital space $a=1,2$ and $a=3,4$. In the lattice problem, the contribution from these potential scatterings tend to compensate the dynamical mass term due to $\hat{H}_W$.

	\end{widetext}

	\section{Derivation of poor man's scaling of Anderson model}\label{Sec:App:Anderson_PMS}
	
	\begin{figure}[t!]
		\includegraphics[width=0.425\textwidth]{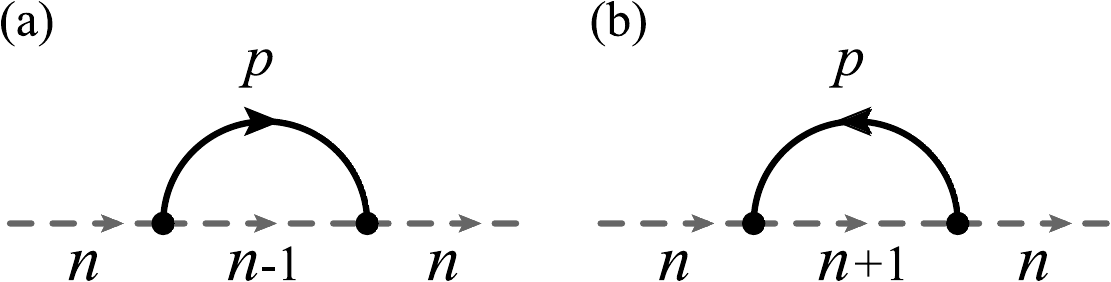}
		\caption{Corrections to energy levels at second-order perturbation theory. (a) Particle contribution. (b) Hole contribution. Dashed lines indicate the $f$ fermion propagators; solid lines indicate $c$ fermion propagators; $p$ is the loop momentum; $n$ indicate the number of $f$ fermions.}
		\label{Fig:T_matrix_And}
	\end{figure}	
	
	The impurity Anderson model is described by $\hat{H}_{\text{And}}=\hat{H}_{c,0}+\hat{H}_{\text{imp},f}+\hat{H}_{\text{imp},cf}+\hat{H}_{\text{imp},J}$ [Eqs.~(\ref{Eq:H_0_c}), (\ref{Eq:H_imp_f}), (\ref{Eq:H_imp_cf}), and (\ref{Eq:H_imp_J})]. The role of the $\hat{H}_{\text{imp},J}$ term is already discussed in the main text. We note several crucial differences in our model from the original Anderson model. First, the electron occupation ranges from $n=0$ to $n=8$ instead of $n=0,1,2$. Second, the density of states of the $c$ fermion band has a strong energy dependence, essentially a linear density of states in the high-energy regime. Despite these complications, we can still apply the poor man's scaling approach following Refs.~\cite{Haldane1978,Cheng2017}.
	
	With a sufficiently large energy cutoff (or $c$ fermion bandwidth), the hybridization term $\hat{H}_{\text{imp},cf}$ induces charge fluctuation in the localized $f$ site, and in turn rise to renormalization of $f$ fermion energy levels as illustrated in Fig.~\ref{Fig:T_matrix_And}. Following Ref.~\cite{Haldane1978}, we compute the flow of $f$ fermion energy levels by integrating out the high energy modes progressively.

	The bare $f$ fermion energy levels are given by $E_n=-\mu_fn+U(n-4)^2/2$. One can easily show that $E_{n+1}+E_{n-1}-2E_n=U$, similar to the symmetry $SU(2)$ Anderson model \cite{Haldane1978}. Upton renormalization, energy levels acquire different corrections. As a result, one needs to examine $E_n$ with $n=0,1,...,8$ altogether. 
	
	Using the second-order perturbation theory [depicted by Fig.~\ref{Fig:T_matrix_And}] and reducing the half bandwidth from $D$ to $D-\delta D$, we derive the expression of $E_n(D-\delta D)$ as follows:
	\begin{align}
		E_n(D-\delta D)=E_n(D)-n\Xi_n^{(p)}(D)-(8-n)\Xi_n^{(h)}(D),
	\end{align}
	where
	\begin{align}
		\Xi_n^{(p)}(D)=&\frac{1}{\mathcal{N}_s}\sum_{\vex{p}}'\sum_{b}\frac{\left|\sum_{a}V^{(\eta)}_{\alpha a}(\vex{p})\Psi^{(\eta)}_{b,a}(\vex{p})\right|^2}{E_{n-1}(D)-E_n(D)+E_b(\vex{p})-\mu_c},\\
		\Xi_n^{(h)}(D)=&\frac{1}{\mathcal{N}_s}\sum_{\vex{p}}''\sum_{b}\frac{\left|\sum_{a}V^{(\eta)}_{\alpha a}(\vex{p})\Psi^{(\eta)}_{b,a}(\vex{p})\right|^2}{E_{n+1}(D)-E_n(D)-E_b(\vex{p})+\mu_c}.
	\end{align}
	In the above expressions, $\Psi_{b,a}(\vex{p})$ and $E_b(\vex{p})$ denotes the wavefunction and energy of the $b$th $c$ fermion band,  
	$\sum_{\vex{p}}'$ sums over $D-\delta D<E_b(\vex{p})-\mu_c<D$ and $\sum_{\vex{p}}''$ sums over $-D<E_b(\vex{p})-\mu_c<-D+\delta D$. The $\Xi_n^{(p)}$ ($\Xi_n^{(h)}$) term denotes the contribution of creating a particle (hole) excitation in $c$ fermion band. The prefactor $n$ in front of $\Xi_n^{(p)}$ indicates the available number of $f$ fermions for transferring to the $c$ fermion conduction bands. Similarly, the prefactor $8-n$ in front of $\Xi_n^{(h)}$ indicates the available occupancy of localized $f$ orbital for accepting fermions from the $c$ fermion valence bands.
	
	With an infinitesimal $\delta D$, one can derive the flow equation
	\begin{align}
		\nonumber\frac{d E_n}{d D}
		=&\frac{n}{\pi}\frac{\sum_b\Delta_b(D+\mu_c)}{E_{n-1}(D)-E_n(D)+D}\\
		\label{Eq:dE_n_dD}&+\frac{8-n}{\pi}\frac{\sum_{b}\Delta_b(-D+\mu_c)}{E_{n+1}(D)-E_n(D)+D}.
	\end{align}
	To analyze the Anderson model, one needs to run the flow equations for all the many-body energy levels.


\end{document}